%% file: ArticleFile_submitArXiv.tex
\documentclass[iop]{emulateapj}
\usepackage{amsmath}
\usepackage{natbib}
\usepackage{hyperref}

\submitted{Accepted to ApJ on November 1, 2017}

\begin{document}

\title{Gemini Spectra of Star Clusters in the Spiral Galaxy M101}
\shorttitle{Star Clusters in M101}

\author{Lesley A. Simanton-Coogan\altaffilmark{1}, Rupali Chandar\altaffilmark{2}, Bryan Miller\altaffilmark{3}, and Bradley C. Whitmore\altaffilmark{4}}

\altaffiltext{1}{Department of Physics, University of North Georgia, Dahlonega, GA 30597}
\altaffiltext{2}{Department of Physics and Astronomy, University of Toledo, Toledo, OH 43606}
\altaffiltext{3}{Gemini Observatory, La Serena, Chile}
\altaffiltext{4}{Space Telescope Science Institute, Baltimore, MD 21218}


\begin{abstract}

We present low resolution, visible light spectra of 41 star clusters in the spiral galaxy M101, taken with the Gemini/GMOS instrument.  We measure Lick indices for each cluster and compare with BaSTI models to estimate their ages and metallicities.  We also measure the line-of-sight velocities.  We find that 25 of the clusters are fairly young massive clusters (YMCs) with ages of hundreds of Myr, and 16 appear to be older, globular clusters (GCs).
There are at least four GCs with best fit ages of $\approx1-3$~Gyr and eight with best fit ages of $\approx5-10$~Gyr.  The mean metallicity of the YMCs is [Fe/H] $\approx-0.1$ and for the GCs is [Fe/H] $\approx-0.9$.  We find a near-continuous spread in both age and metallicity for our sample, which may indicate that M101 had a more-or-less continuous history of cluster and star formation.  From the kinematics, we find that the YMCs rotate with the HI gas fairly well, while the GCs do not.  We cannot definitively say whether the GCs sampled here lie in an inner halo, thick disk, or bulge/psuedobulge component, although given the very small bulge in M101, the last seems unlikely.  The kinematics and ages of the YMCs suggest that M101 may have undergone heating of its disk or possibly a continuous merger/accretion history for the galaxy.

\end{abstract}

\section{Introduction}

Star clusters have formed with a wide range of ages from millions of years to $\sim12$ billion years, making them excellent tracers of the histories of their host galaxies.  The optical colors of clusters older than $\sim1$~Gyr suffer from the well-known age-metallicity degeneracy.  Spectra contain absorption lines of specific elements whose strength is dominantly dependent upon either age or metallicity, but not both, allowing the age-metallicity degeneracy to be broken.  Here, we use spectra obtained with the Gemini/GMOS instrument of bright star clusters in the spiral galaxy M101 to determine their ages and metallicities.

The ages and metallicities of this sample of star clusters can shed light on the cluster populations of M101 and its history.  In the Milky Way, the globular clusters (GCs) have a bimodal metallicity distribution that correlates with their spatial distributions.
The high metallicity GCs (with a peak [Fe/H]$\sim-0.5$) are concentrated towards the center of the galaxy, and believed to have formed as part of the bulge, while the low metallicity GCs (peak [Fe/H]$\sim-1.5$) have a more extended spherical distribution, and are part of the halo.
More specifically, the range of metallicities observed for bulge clusters ($-1.0 \leq$~[Fe/H]~$\leq +0.5$, \citet{min95_2}) suggests that the MW bulge has been built at least partially by mergers \citep{min95}.  The metal-poor outer halo GCs in the MW have a spread in ages ($\sim 10.5-13$~Gyr, \citet{lea13}) and no metallicity gradient, which also supports a scenario where the accretion of satellite galaxies was an important process in building up this component \citep{sea78}.  Intermediate age clusters ($\sim 6-8$~Gyr) in the halos of M31 and M33 (a bulgeless spiral similar to, but less massive than, M101) are also believed to have an accretion origin \citep{bro09}.

We can also use the kinematics of a population of star clusters to disentangle the components and structure of their host galaxies.  In the MW, old GCs are found in the bulge and halo components while younger, open clusters are found in the thin or thick disks \citep{por10}.  In M33, \citet{cha02} find a population of old GCs associated with the disk/pseudobulge and old GCs in the halo as well as a population of intermediate age clusters with evidence of disk motions.  

We use velocities measured from Gemini/GMOS spectra of a sample of M101 YMCs and GCs to disentangle the structure of the galaxy, and add insight into the formation history.  Both the line-of-sight velocities as well as the rotational velocity, $v_{\text{rot}}$, can reveal the structure of a cluster population.  For nearly face-on galaxies such as M101, high dispersion in the line-of-sight velocities ($\sigma$) are indicative of a spheroidal component such as a bulge or halo, while low $\sigma$ correspond to disk populations.  In the MW, old open clusters have $\sigma_{\text{old, open}}=28$~km/s \citep{sco95}, disk/bulge GCs have $\sigma_{\text{bulge,GCs}}=67$~km/s \citep{cot99}, and halo GCs have $\sigma_{\text{halo,GCs}}=114$~km/s \citep{zin85}.  The ratio $v_{\text{rot}}/\sigma$ is $>1$ for rotationally supported systems (e.g. disks) while $v_{\text{rot}}/\sigma<1$ indicates a pressure-supported, spheroidal system (e.g. bulge, halo).  

\citet{kor10} use $HST$ photometry and high resolution ($R\equiv\lambda/\text{FWHM}\simeq15,000$) spectra from the Hobby-Eberly Telescope to decompose and determine velocity dispersions of the nuclear regions of several late-type spiral galaxies, including M101.  They find the inner velocity dispersion (most likely corresponding to the pseudobulge) to be low, $\sigma_{\text{pseudobulge}}=27\pm4$~km/s, which may mean it is more of an ``inner disk'' than a bulge in structure.  \citet{van14} examined the surface brightness profile of M101 out to $R_{\text{gc}}=70$~kpc in order to fit the bulge, disk, and halo components.  They find a surprisingly low halo mass fraction $f_{\text{halo}}=M_{\text{halo}}/M_{\text{tot}}=0.003_{-0.003}^{+0.006}$ compared to that of the MW $f_{\text{halo}}\approx0.02$.  In this paper, we see how properties of clusters in M101 fit into this highly disk dominated galaxy.

This paper is arranged as follows:  We describe the cluster candidate selection and $Gemini/GMOS$ observations in \S\ref{sec:Observations}.  In \S\ref{sec:Measurements}, we measure line-of-sight velocities for our clusters as well as Lick indices, which we use to determine the cluster ages and metallicities.  In \S\ref{sec:Results}, we compare the velocities to the cluster spatial distribution and ages and examine the cluster age, metallicity, and spatial distributions.  We discuss the results in \S\ref{sec:Discussion}.  Finally in \S\ref{sec:Conclusions}, we list our conclusions.

\section{Observations}
\label{sec:Observations}

\subsection{Selecting Clusters for Spectroscopic Observations}
\label{subsec:CandidateObservations}

Candidate star clusters were identified from $BVI$ images taken with the Advanced Camera for Surveys/ Wide Field Channel instrument on board the $Hubble$ $Space$ $Telescope$ in November 2002 as part of Program GO 9490 (PI:  K. Kuntz)\footnote{Based on observations made with the NASA/ESA Hubble Space Telescope, and obtained from the Hubble Legacy Archive, which is a collaboration between the Space Telescope Science Institute (STScI/NASA), the Space Telescope European Coordinating Facility (ST-ECF/ESA) and the Canadian Astronomy Data Centre (CADC/NRC/CSA).}.  The methods were similar to those described in \citet{sim15}, which we summarize here.  

We detect $\sim383,000$ sources which include star clusters, bright individual stars, and background galaxies, and measure their brightnesses within circular apertures out to 5 pixels with background estimation within annuli from 7 to 13 pixels.  We apply aperture corrections \citep{sir05} and zero points to get results in the VEGA-MAG system \citep{boh07,mac07}\footnote{\url{http://www.stsci.edu/hst/acs/analysis/zeropoints}}.  We select clusters using the following criteria:

\begin{itemize}
\item Brighter than $m_V$ of 21.5, to ensure high S/N.
\item Concentration index (CI) $> 1.15$, where CI is the difference between $m_V$ measured within 1 pixel and 3 pixel apertures to eliminate point sources; point sources have CI values that peak around $1.00$ with a standard deviation of $0.06$.
\item We prioritize clusters with colors similar to those of Galactic GCs, i.e. $0.55 < B-V < 2.0$ and $0.75 < V-I < 2.5$ to help exclude background galaxies while ensuring that old GC candidates are included.
\end{itemize}

\begin{figure}[htp]
\begin{center}
\includegraphics[width=\columnwidth]{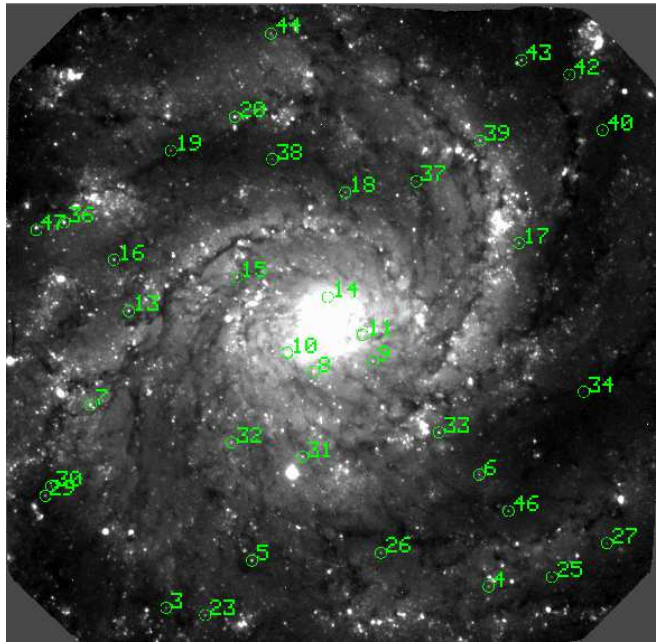}
\caption[Gemini-North Images of GMOS Mask 1]{g-band image of M101 taken with Gemini-North showing the objects observed in mask 1. \label{mask1}}
\end{center}
\end{figure}

\begin{figure}[htp]
\begin{center}
\includegraphics[width=\columnwidth]{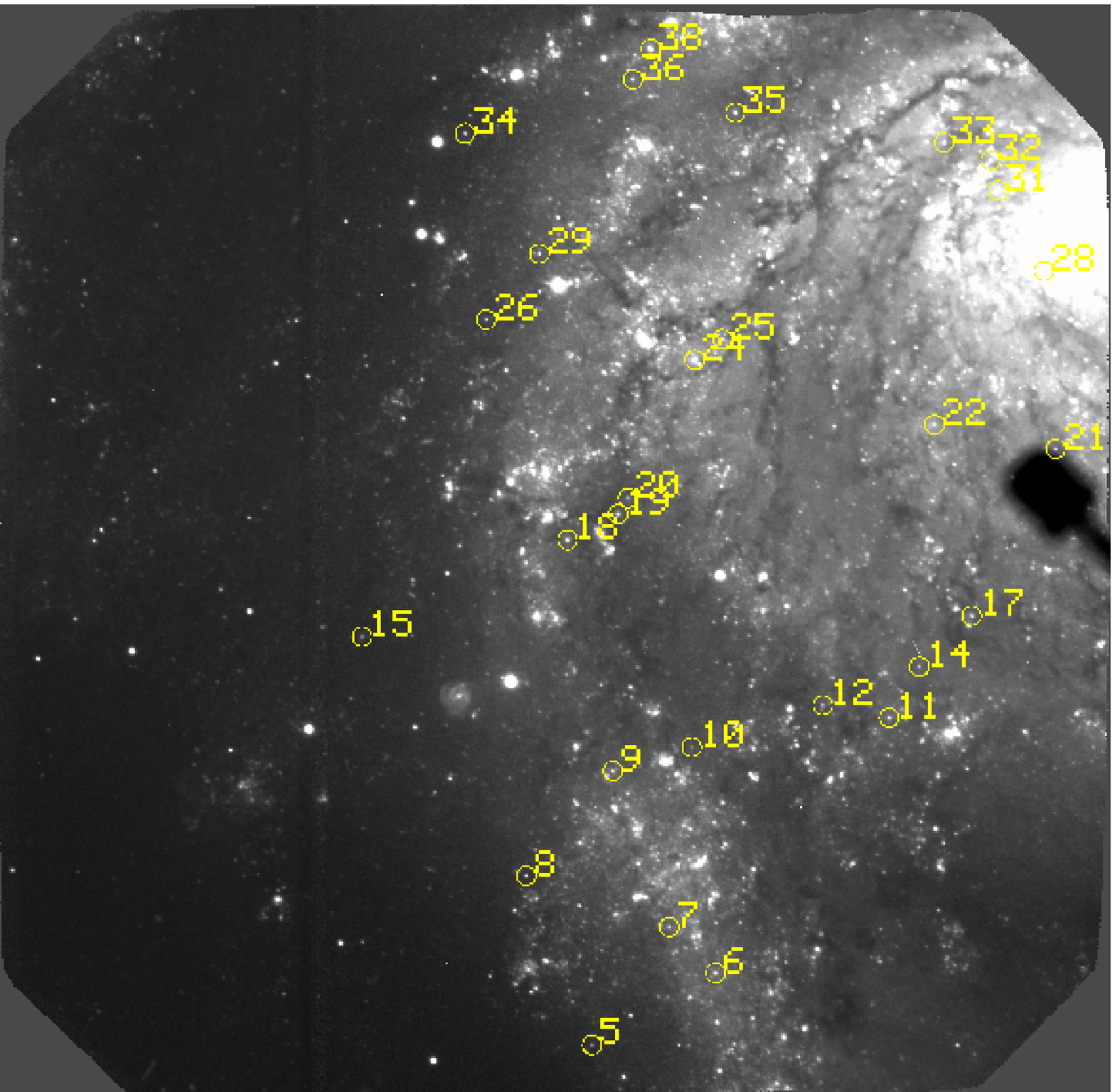}
\caption[Gemini-North Images of GMOS Mask 2]{g-band image of M101 taken with Gemini-North showing the objects observed in mask 2. \label{mask2}}
\end{center}
\end{figure}

\begin{figure}[htp]
\begin{center}
\includegraphics[width=\columnwidth]{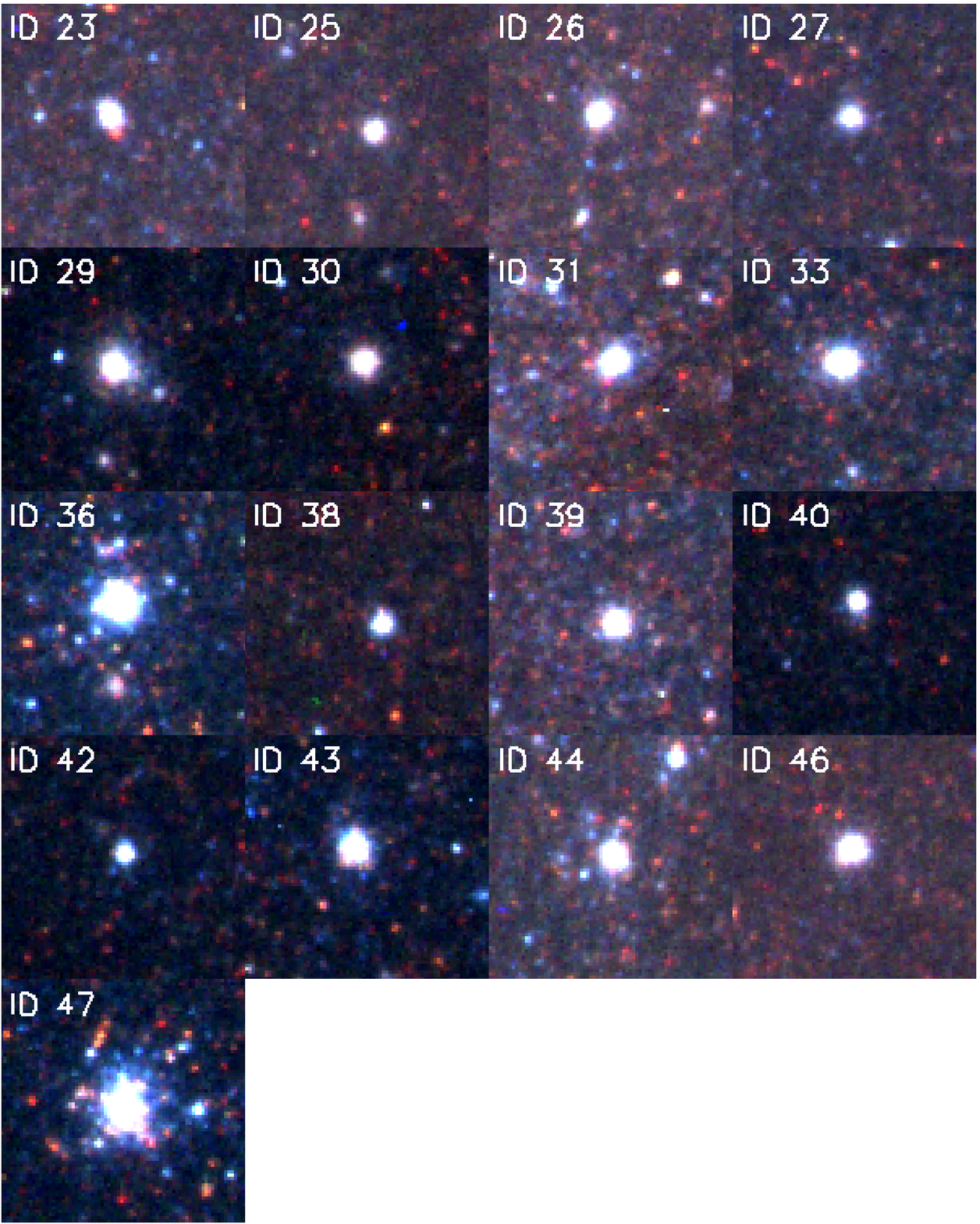}
\includegraphics[width=\columnwidth]{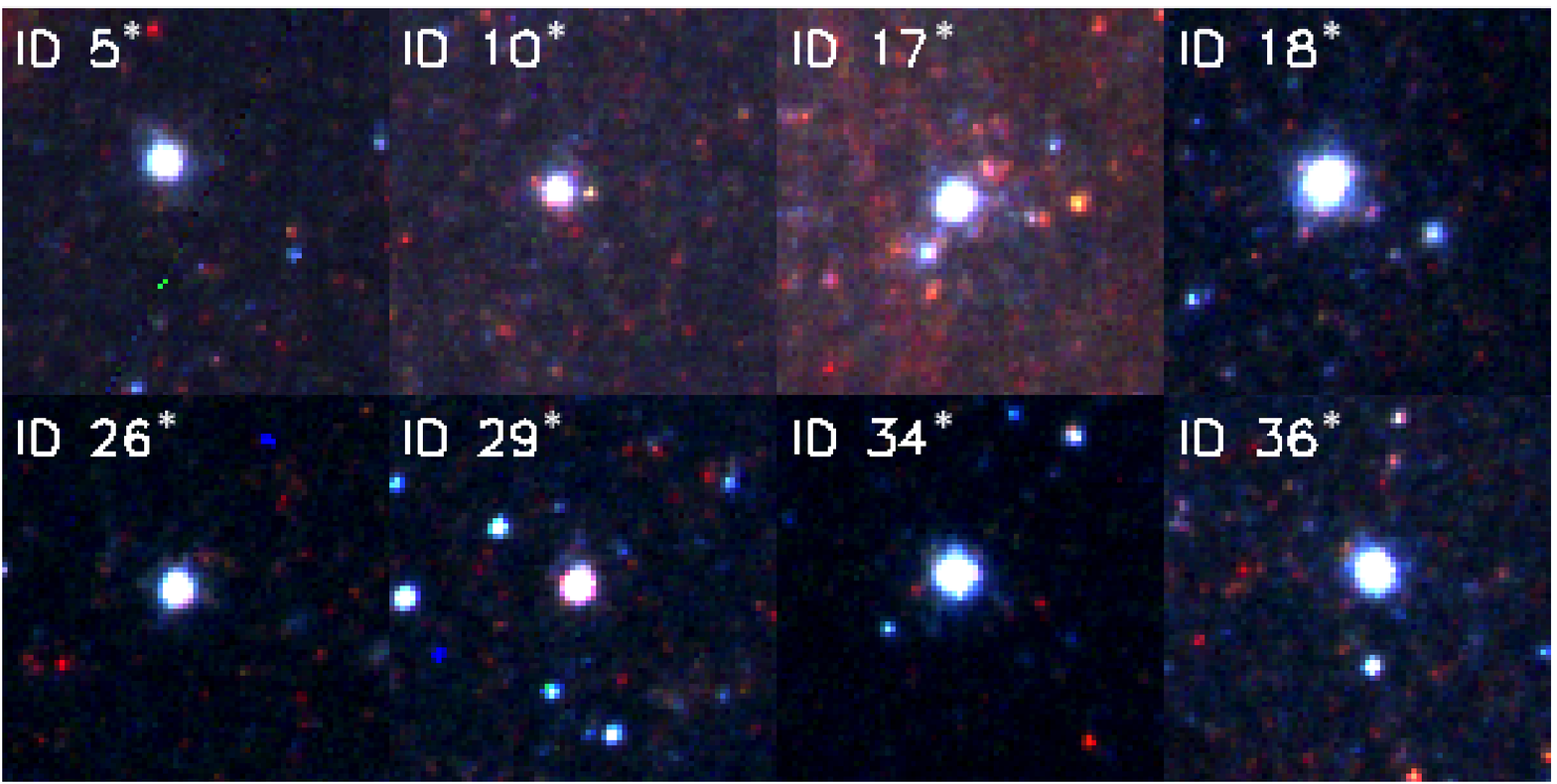}
\caption[$HST$ $BVI$ Color Images of M101 YMCs with Spectra]{$BVI$ color images from the $HST$ of the YMCs for which we have obtained Gemini spectroscopy.  Each cutout is approximately $7.35''$ $\times$ $7.35''$. The top portion shows clusters located on mask 1 of the GMOS data, and the bottom set shows clusters located on mask 2.  Asterisks denote mask 2 clusters.  \label{stamps_YMCs}}
\end{center}
\end{figure}

\begin{figure}[htp]
\begin{center}
\includegraphics[width=\columnwidth]{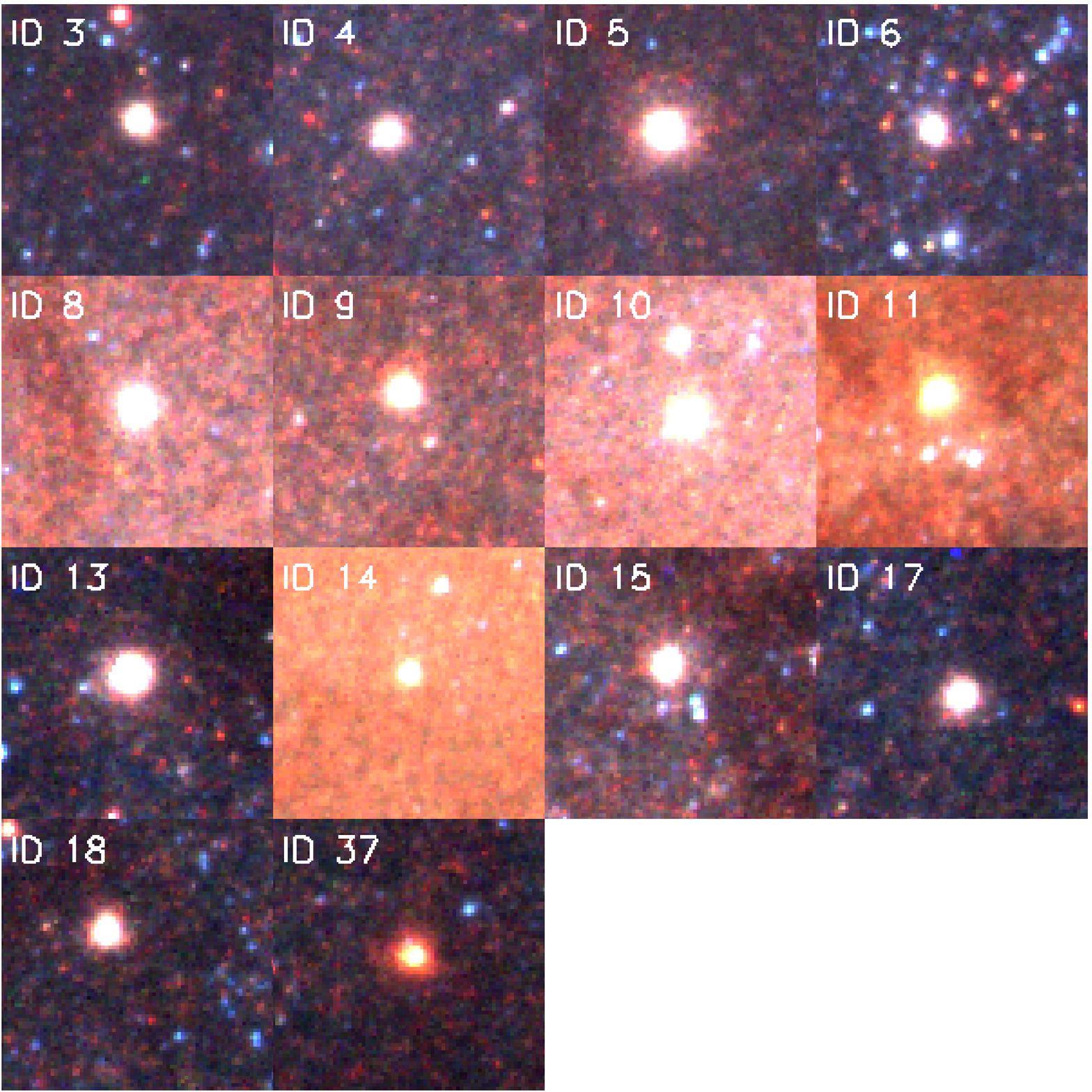}
\includegraphics[width=0.5\columnwidth]{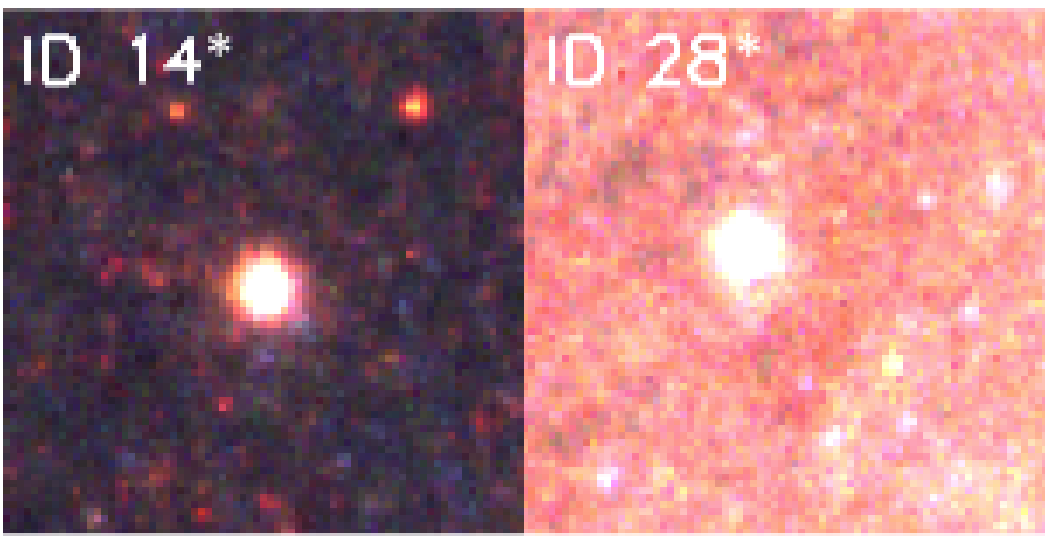}
\caption[$HST$ $BVI$ Color Images of M101 GCs with Spectra]{$BVI$ color images from the $HST$ of the GCs for which we have obtained Gemini spectroscopy.  Each cutout is approximately $7.35''$ $\times$ $7.35''$. The top portion shows clusters located on mask 1 of the GMOS data, and the bottom set shows clusters located on mask 2.  Asterisks denote mask 2 clusters.  \label{stamps_GCs}}
\end{center}
\end{figure}

We custom designed two multi-slit masks in order to observe as many clusters as possible.  These are shown in Figures~\ref{mask1} and \ref{mask2} and contain 55 cluster candidates, 23 of which have red colors typical of Galactic GCs.  Color postage stamp images taken with $HST$ in the $BVI$ filters showing all clusters for which spectra were obtained are shown in Figures~\ref{stamps_YMCs} and \ref{stamps_GCs} and spatial and photometric properties of our spectroscopic targets are listed in Table~\ref{photTable}.

\vfill
\vfill

\vfill
\vfill

\vfill
\vfill

\vfill
\vfill

\input{table1}

\subsection{Gemini/GMOS Observations}
\label{subsec:GemObservations}

The spectroscopic observations were taken with the GMOS-North instrument (GMOS-N) over several nights in the 2008A semester as part of Gemini program GN-2008A-Q-55.  Seeing varied between $0.61''$ and $1.22''$.  The slow read mode was utilized with the B600-G5303 grating and 4 $\times$ 2 binning (dispersion $\times$ spatial).  The slit widths were fixed at $1.0''$, and the slit lengths were variable to accommodate as many objects as possible on each mask.  Figures~\ref{mask1} and \ref{mask2} show locations of objects observed in each mask.  Labels show the ID numbers used to refer to each object in this paper.  There were 10 exposures of mask 1 and four exposures of mask 2, each with an exposure time of 3600~s.  

The spectra are reduced using development versions of the Gemini IRAF package along with additional steps coded in PyRAF and IDL. A brief outline of the reduction steps are given below; for a more detailed description, see Appendix A2 of Trancho et al. (2007). We combine the bias frames for each night and subtract from each target frame. We use the MOSPROC pipeline script to apply the flat fielding, wavelength calibration, cosmic ray cleaning, quantum efficiency correction, and background subtraction steps and finally extract the one-dimensional spectra. The best wavelength calibration fits a sixth order spline3 function along the spatial axis (median RMS values $\sim 0.43 \pm 0.07$). Most of the custom steps in this process, including the quantum efficiency corrections, are now included in the Gemini IRAF package since v1.13\footnote{\url{http://www.gemini.edu/sciops/data-and-results/processing-software}}.

Each MOS exposure is taken at a different orientation with respect to the parallactic angle and therefore has unique slit losses due to atmospheric refraction. The slit losses are corrected using the IDL script SLITCORR\footnote{\url{http://drforum.gemini.edu/topic/slitcorr-slit-loss-corrections}} that applies the method of \citet{fil82}. The parameters of the correction, especially the position within the slit, are tweaked in order to make the continuum shapes as uniform as possible.  Changes to each spectrum are multiplicative to correct for light losses and do not apply shifts in the dispersion direction and so should have no impact on either the measurement of velocities (see \S\ref{subsec:MeasureVelocities}) or the wavelength regions used in measuring Lick indices (see \S\ref{subsec:MeasureAgeMetals}).  

The final reduction steps include averaging the spectra of each target and applying relative flux calibration. The refraction-corrected spectra of each object are shifted to a common heliocentric velocity correction before they are averaged together by weighting by the mode of each spectrum.  The averaged spectra are inspected and any remaining bad pixels as well as spurious lines from incomplete background subtraction of sky lines are masked and then cleaned. Finally the relative flux calibration, correction for the total system throughput, is applied using baseline calibration observations of the spectrophotometric standard HZ44.  The spectra do not have absolute flux calibration.

\begin{figure}[htp]
\begin{center}
\includegraphics[width=\columnwidth]{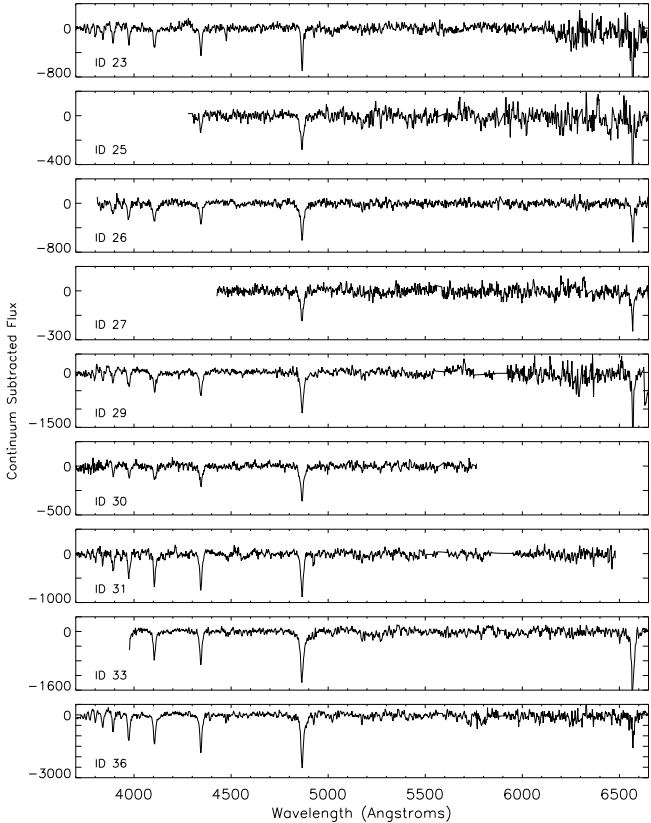}
\caption[GMOS Spectra of Mask 1 YMCs]{Gemini/GMOS spectra of the ``young" massive clusters on mask 1.  All spectra shown here through Fig.~\ref{GCspec1_mask2} have been continuum subtracted.  \label{YMCspec1_mask1}}
\end{center}
\end{figure}

\begin{figure}[htp]
\begin{center}
\includegraphics[width=\columnwidth]{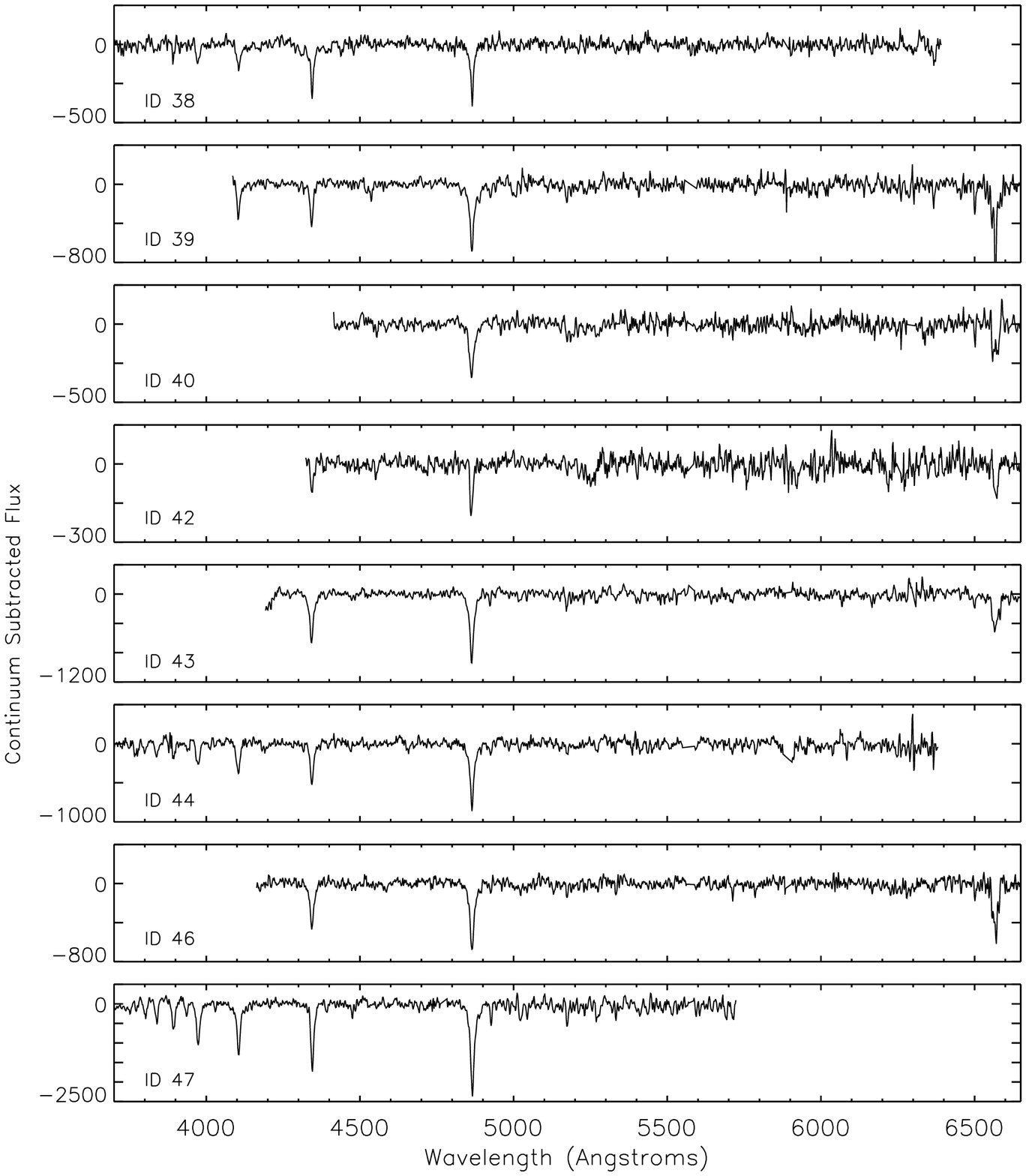}
\caption[GMOS Spectra of Mask 1 YMCs Continued]{Additional GMOS spectra of the YMCs on mask 1.  \label{YMCspec2_mask1}}
\end{center}
\end{figure}

\begin{figure}[htp]
\begin{center}
\includegraphics[width=\columnwidth]{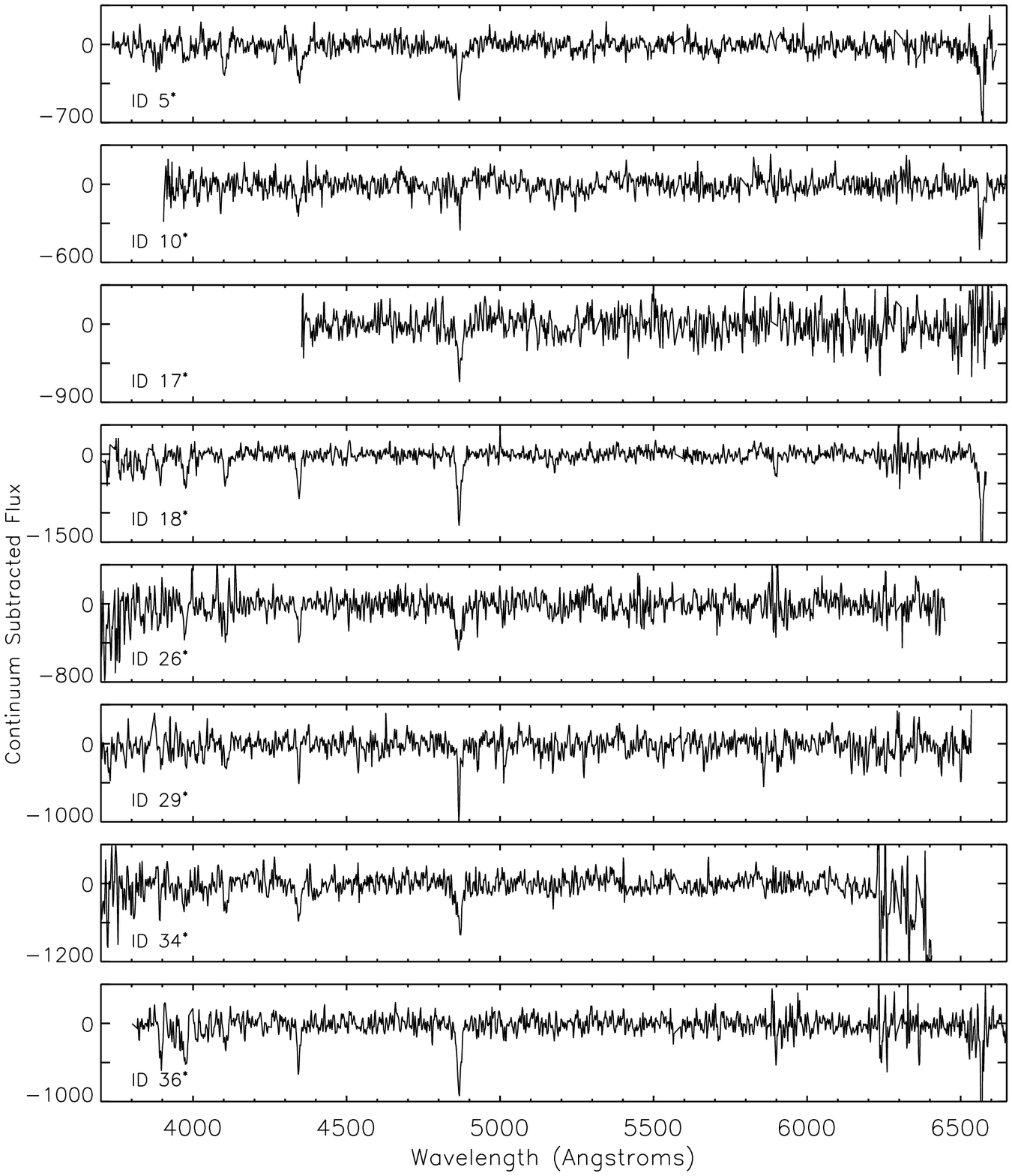}
\caption[GMOS Spectra of Mask 2 YMCs]{GMOS spectra of the YMCs on mask 2.  \label{YMCspec1_mask2}}
\end{center}
\end{figure}

\begin{figure}[htp]
\begin{center}
\includegraphics[width=\columnwidth]{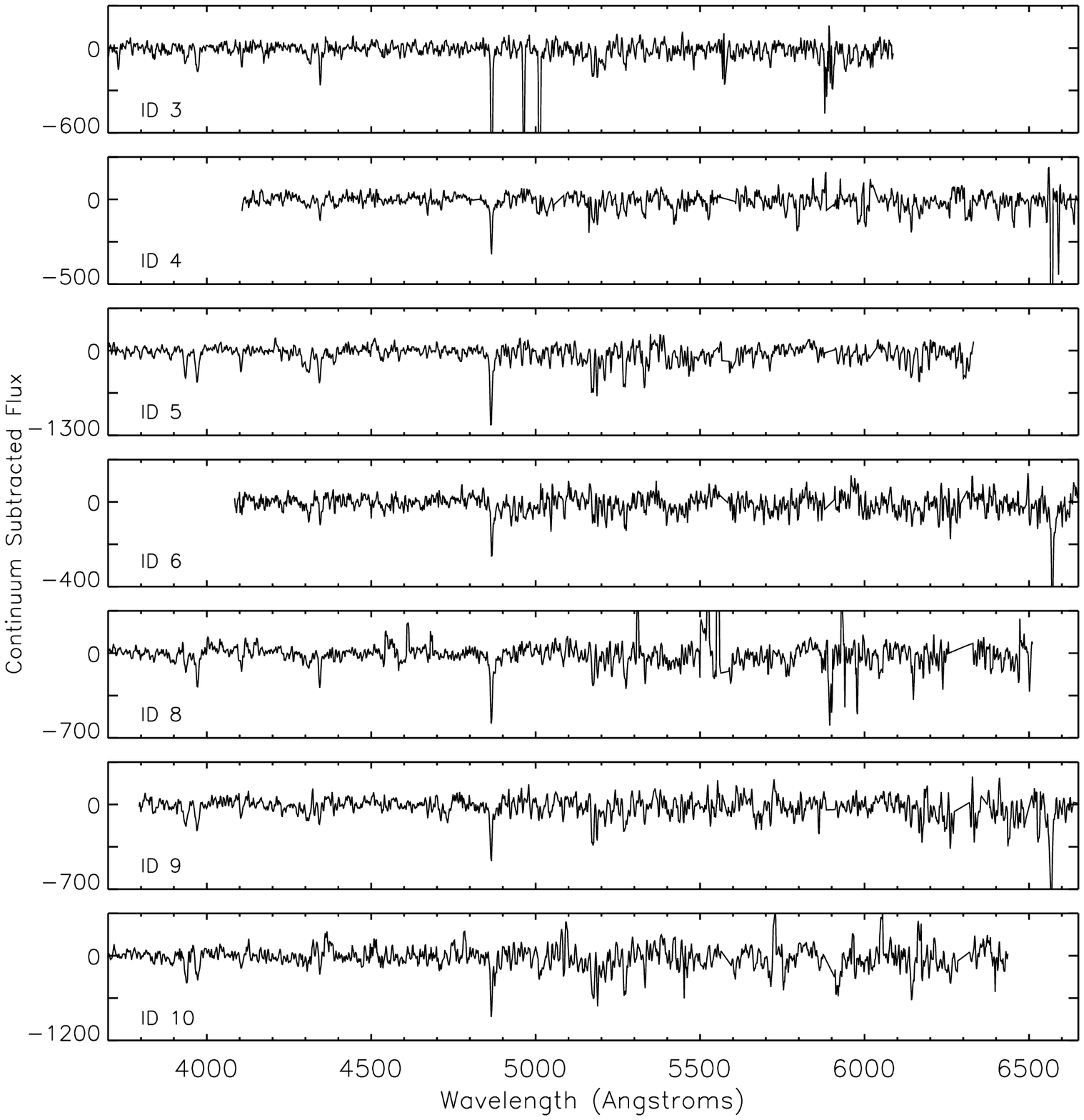}
\caption[GMOS Spectra of Mask 1 GCs]{GMOS spectra of the GCs on mask 1.  \label{GCspec1_mask1}}
\end{center}
\end{figure}

\begin{figure}[htp]
\begin{center}
\includegraphics[width=\columnwidth]{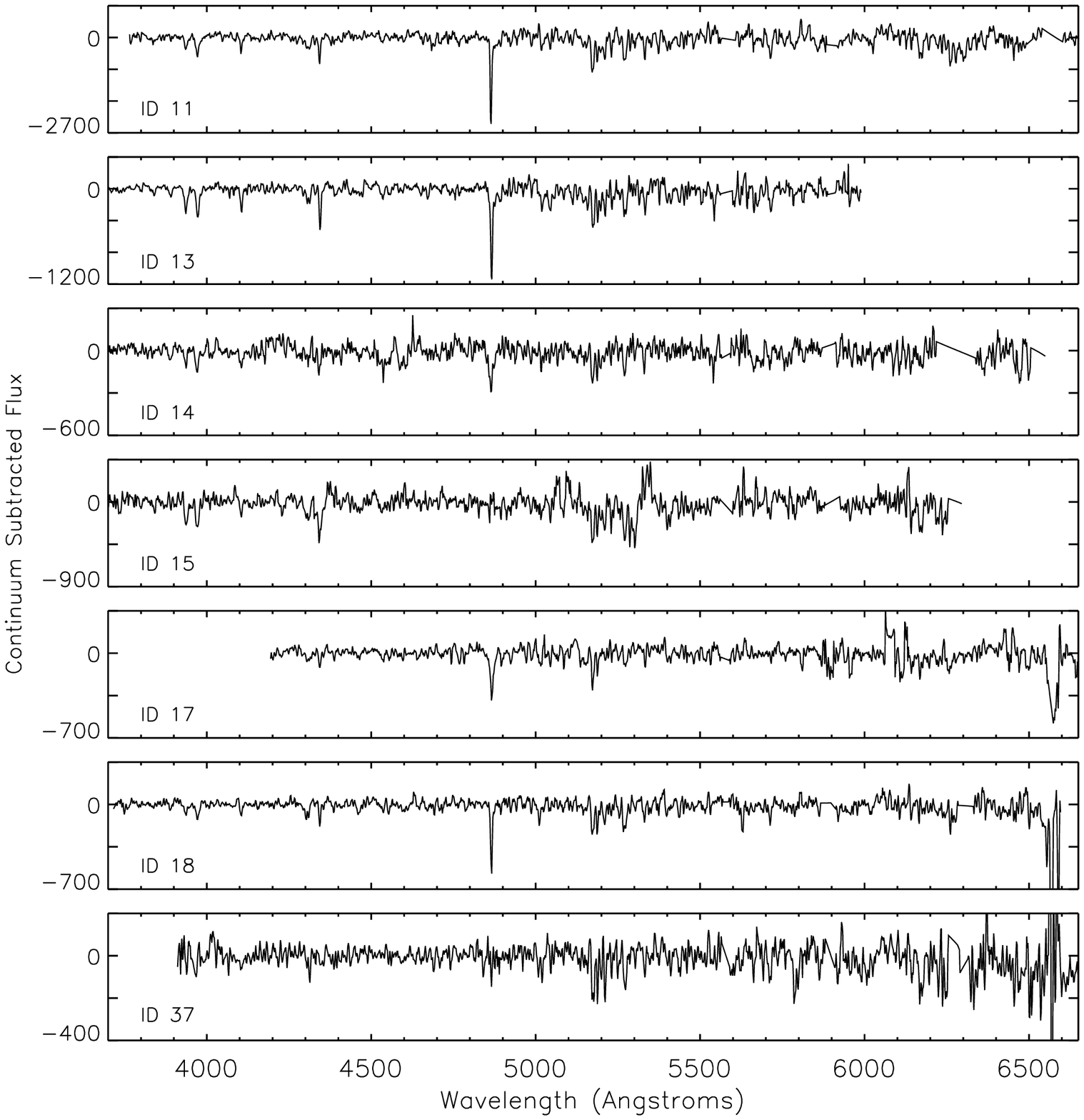}
\caption[GMOS Spectra of Mask 1 GCs Continued]{Additional GMOS spectra of the GCs on mask 1.  \label{GCspec2_mask1}}
\end{center}
\end{figure}

\begin{figure}[htp]
\begin{center}
\includegraphics[width=\columnwidth]{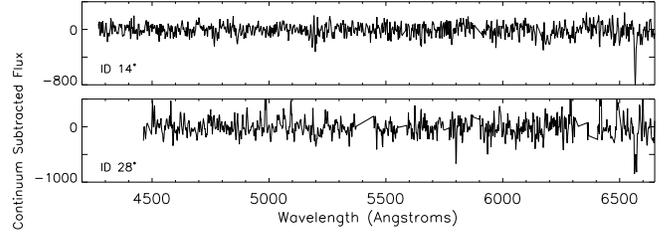}
\caption[GMOS Spectra of Mask 2 GCs]{GMOS spectra of the GCs on mask 2.  \label{GCspec1_mask2}}
\end{center}
\end{figure}

Figures~\ref{YMCspec1_mask1} through \ref{GCspec1_mask2} show the final continuum subtracted spectra.  We have grouped the clusters into preliminary ``young" and ``old" categories based on a visual inspection of the spectral features with ``young" massive cluster (YMC) spectra containing strong Balmer lines (equivalent width of the H beta line, $W_{\text{H}\beta} \gtrsim 5.34$~\AA) and ``old" GC spectra containing weak Balmer and strong metal lines (Ca H and K at $\sim$3900~\AA, G band at $\sim$4300~\AA, and Fe 5170).  Table~\ref{specTable} lists properties of the spectra such as their full wavelength range, S/N ratio, and $W_{\text{H}\beta}$. 

14 objects on the masks are not included in our final cluster sample.  They include five background galaxies (identified from redshifted Ca H and K lines), two red (super)giant stars in the Galaxy, and six clusters showing emission lines (because they are close to H II regions and probably quite young, $<10$~Myr).  This leaves 25 YMCs and 16 GCs.

\vfill
\vfill

\vfill
\vfill

\vfill
\vfill

\vfill
\vfill

\vfill
\vfill

\vfill
\vfill

\input{table2}

\section{Measurements}
\label{sec:Measurements}

\subsection{Velocity Measurements}
\label{subsec:MeasureVelocities}

We determine the radial velocity of each cluster by comparing its spectrum with a template spectrum.  We use the PyRaf task FXCOR which performs a Fourier cross-correlation of the two spectra, and in each case there is a strong peak at the velocity of the cluster.  The template spectra are high S/N clusters from within our own sample so that the velocities measured for each object are robust on a relative scale.  We choose one YMC (ID 36 on mask 1, see Fig.~\ref{YMCspec1_mask1}) and one GC (ID 11, see Fig.~\ref{GCspec2_mask1}) as templates for measuring the relative velocities for the YMCs and GCs respectively.  By using one of our own clusters as a template, we ensure that the template spectra are a decent match to the object spectra, and remove any complications from mismatches in resolution and other subtle effects that can arise if the template spectra come from models or are taken with a different instrument or in a separate program.

Both the object and template spectra for each FXCOR run were first continuum subtracted and various wavelength ranges were tested.  The wavelength ranges chosen for each object were those with the largest range that contained good S/N, clear lines, and did not have masked regions.  

For each template spectrum, we calculate velocities using Doppler Shift, $v = c(\frac{\lambda_{\text{cluster}}-\lambda_{\text{rest}}}{\lambda_{\text{rest}}})$, from the H$\beta$, H$\gamma$, and H$\delta$ absorption lines and take the mean as the velocity of the template.  The appropriate template velocity is added to the relative velocity of each cluster to estimate an ``absolute'' velocity, $v_{\text{cluster}}$, for each cluster.  These velocities are reported in Table~\ref{kinTable}.  

We use two separate approaches to estimate the uncertainties on the velocity determinations.  Velocity errors from FXCOR are computed from the ratio of the height of the strongest peak to the height of the average peak in the correlation (i.e. the anti-symmetric noise of the correlation function) according to the methods of \citet{ton79}.  Note that the width of the peak combined with the width of the template reveal the velocity dispersion of the object, which is only relevant for galaxy spectra as velocity dispersions within star clusters are quite low ($\sim2-10$~km/s).  We do not measure the internal velocity dispersions of the clusters here.

For our second method, we estimate velocity errors as the standard deviation of velocities of individual exposure spectra that make up each combined cluster spectrum ($\sigma_{v_{\text{cluster}}}$, see Table~\ref{kinTable}).  We find these $v_{\text{cluster}}$ errors to be more comprehensive and realistic than the FXCOR calculated errors as they take into account the variations between exposures with different random noise patterns or which were taken on different nights.  We find that the median $\sigma_{v_{\text{cluster}}}$ ($\sim32$~km/s for GCs and $\sim40$~km/s for YMCs) match our expectation that the YMCs should have higher overall errors due to broader Balmer lines with less sharply defined peaks, while the median FXCOR errors do not ($\sim19$~km/s for GCs and $\sim11$~km/s for YMCs).

\input{table3}

\subsection{Measuring Ages and Metallicities Using BaSTI Models}
\label{subsec:MeasureAgeMetals}

We estimate the age and metallicity of each cluster by comparing the strengths of absorption lines measured from the cluster spectra to those of synthetic spectra of SSPs from the Bag of Stellar Tracks and Isochrones (BaSTI) library\footnote{Available at \url{http://basti.oa-teramo.inaf.it}.} \citep{per09}.  We chose the BaSTI models because they incorporate the full evolution of thermally pulsing AGB stars (important at ages of a few 100 Myr), make predictions for both scaled-solar and $\alpha$-enhanced abundances, and provide high resolution optical spectra of single stellar populations.

The spectral indices are an equivalent width, given by $W_{\lambda} = \int_{\lambda_1}^{\lambda_2} (1-\frac{F_\lambda}{F_C})d\lambda$ where $\lambda_1$ and $\lambda_2$ define the region where the line is measured (the passband), $F_\lambda$ is the flux of the absorption line and $F_C$ is the flux of the continuum.  Here, the indices are measured with respect to a pseudo continuum drawn between regions redward and blueward of the passband.  We use the index system defined from several decades of study of stellar, cluster, and galaxy spectra taken on the $Lick$ $Observatory$/image dissector scanner (Lick/IDS indices) \citep{fab85, bur86, wor94, wor97}.  

We first shift the cluster spectra by the doppler shift of the Balmer line closest to each passband.  We then measure Lick/IDS indices of both the cluster spectra and BaSTI library of synthetic spectra using the LICK\_EW IDL function\footnote{This software was written by Genevieve Graves and is part of the EZ\_AGES IDL code package described in \citet{gra08} and available at \url{http://astro.berkeley.edu/~graves/ez_ages.html}.}, which smooths the input spectrum to the $\sim8-11$~\AA$pixel^{-1}$ resolution of the Lick/IDS index system and determines the $W_{\lambda}$ of each index in the smoothed spectrum.  \citet{puz14} compared observations of standard stars taken by Gemini/GMOS and Lick/IDS and found corrections (of the form $W_{\lambda, Lick} = W_{\lambda, GMOS} - \delta$, where $\delta$ are constants) are required to convert spectral indices measured from Gemini spectra to the Lick system.  We apply these calibration constants to the $W_{\lambda}$ measured from the cluster spectra, using values from \citet{puz14} for the closest instrument set-up available: Gemini/GMOS-N, B600 grism with $0.75''$ slit widths and 2 $\times$ 2 binning.

\begin{figure}[htp]
\begin{center}
\includegraphics[width=0.49\columnwidth]{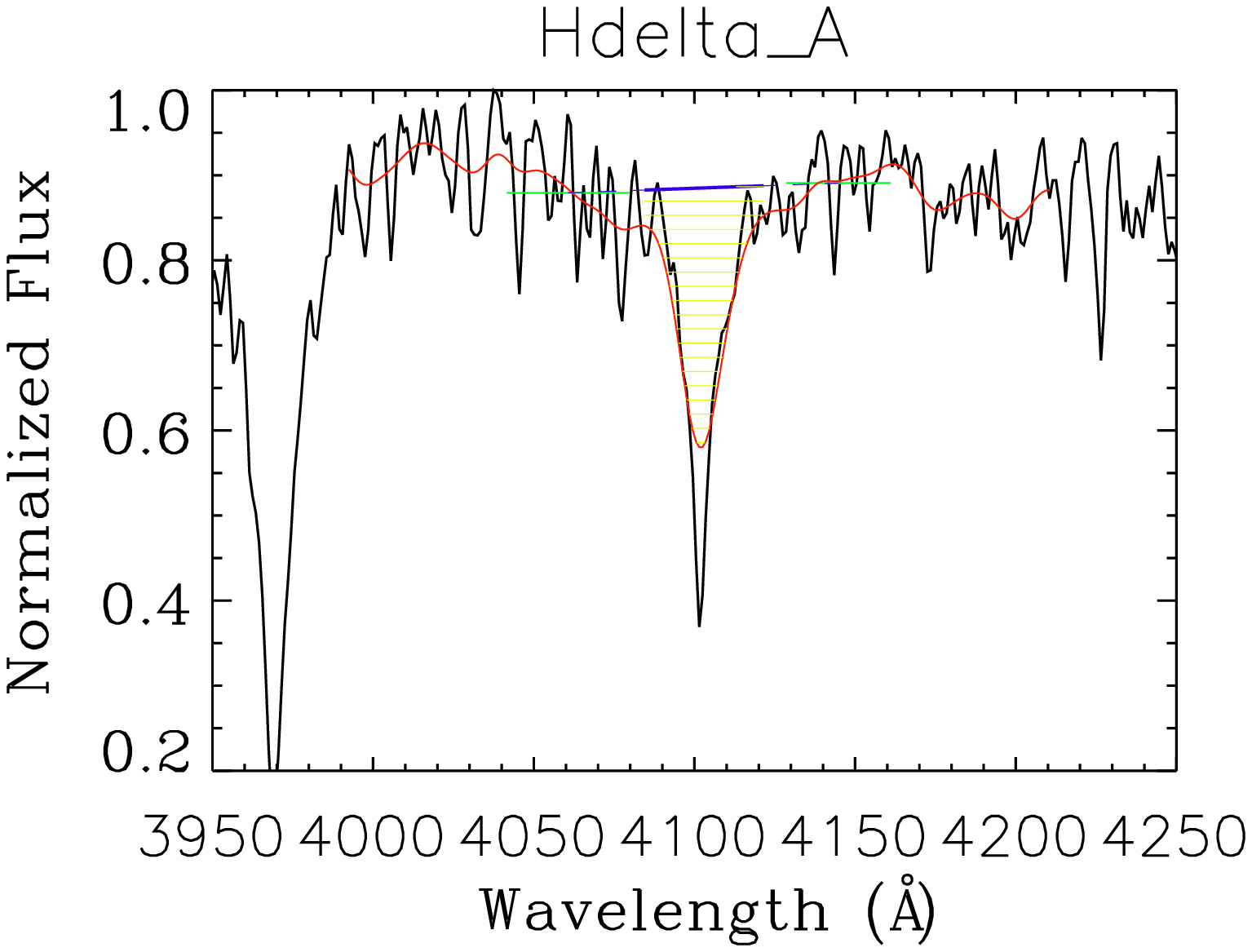}
\includegraphics[width=0.49\columnwidth]{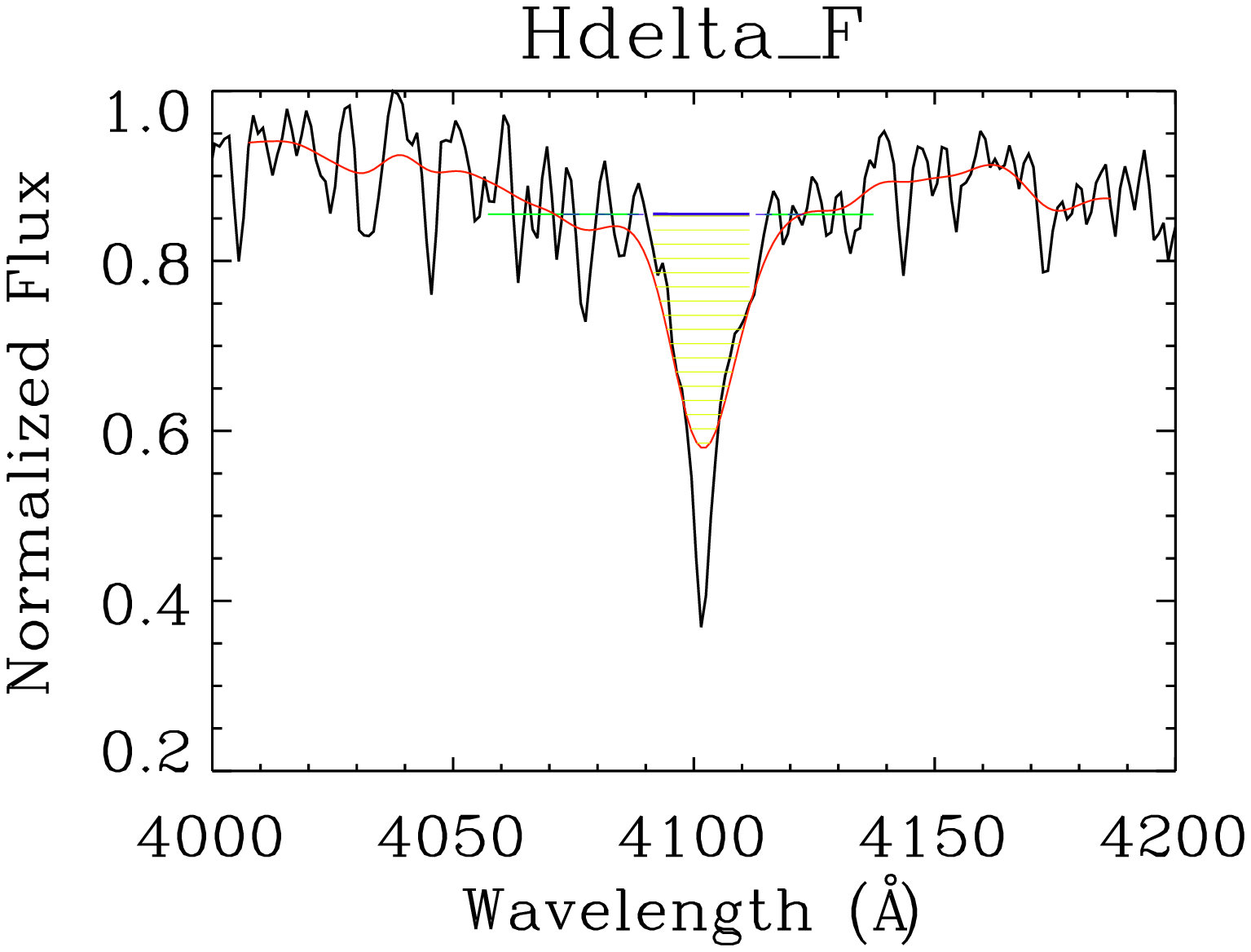}
\includegraphics[width=0.49\columnwidth]{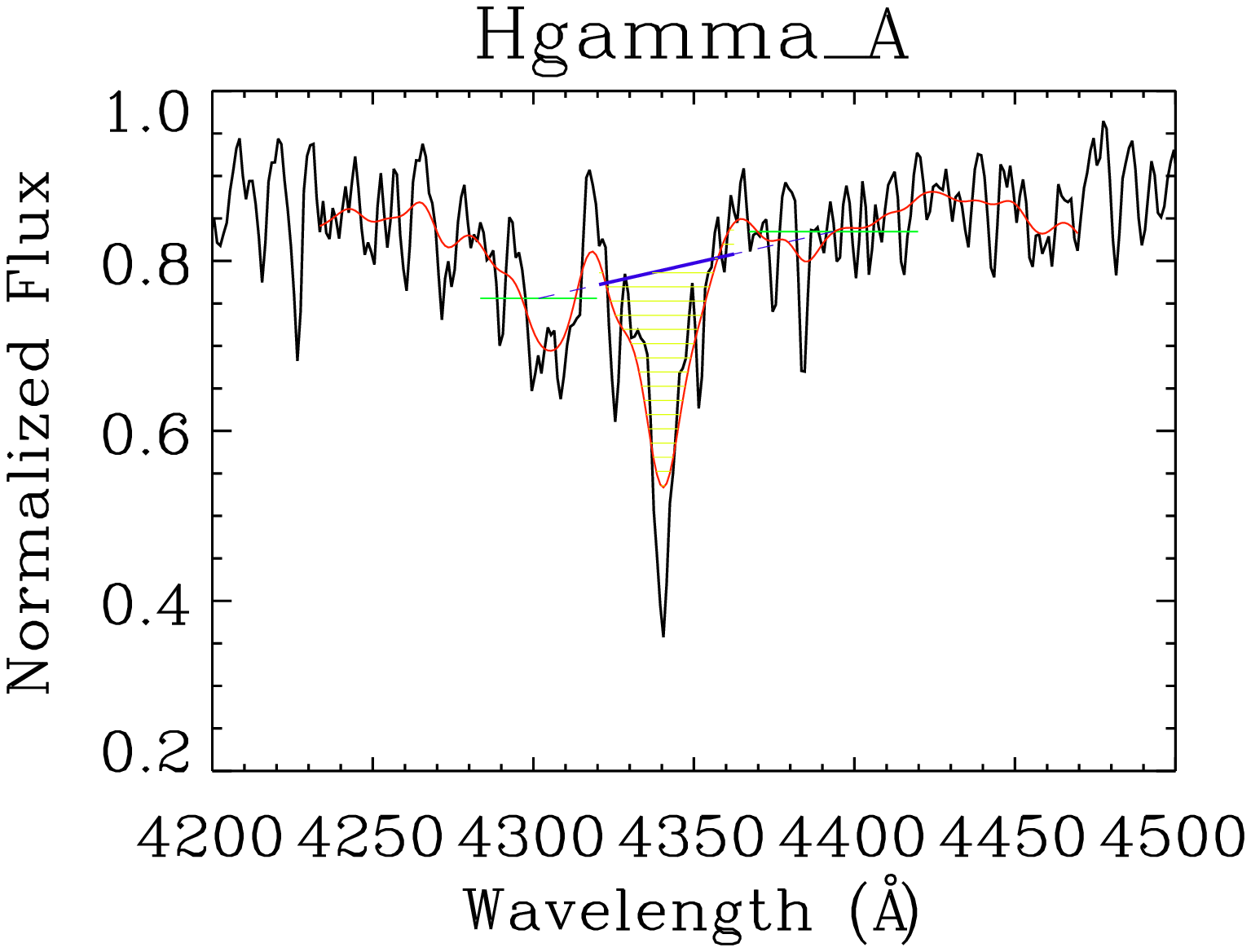}
\includegraphics[width=0.49\columnwidth]{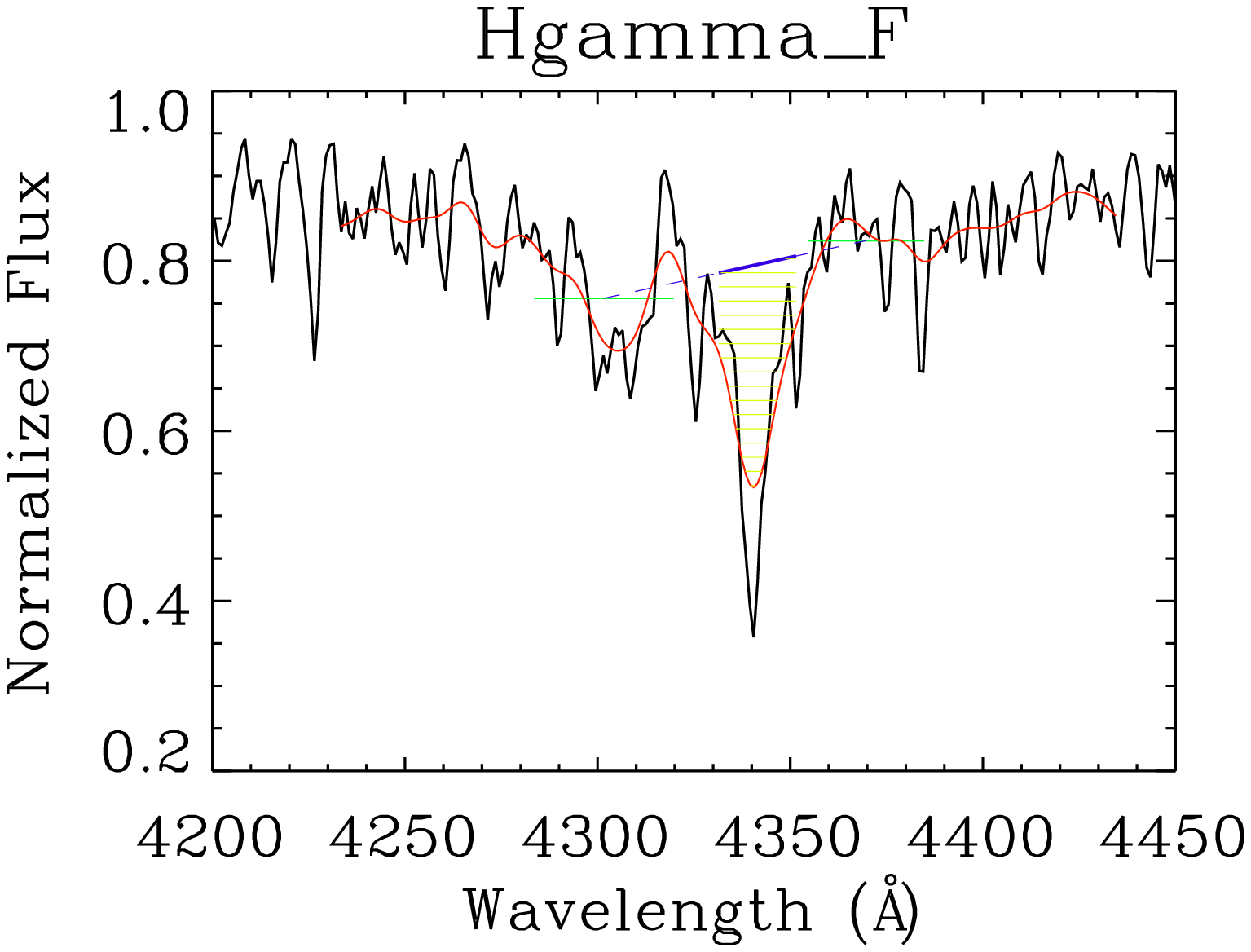}
\includegraphics[width=0.49\columnwidth]{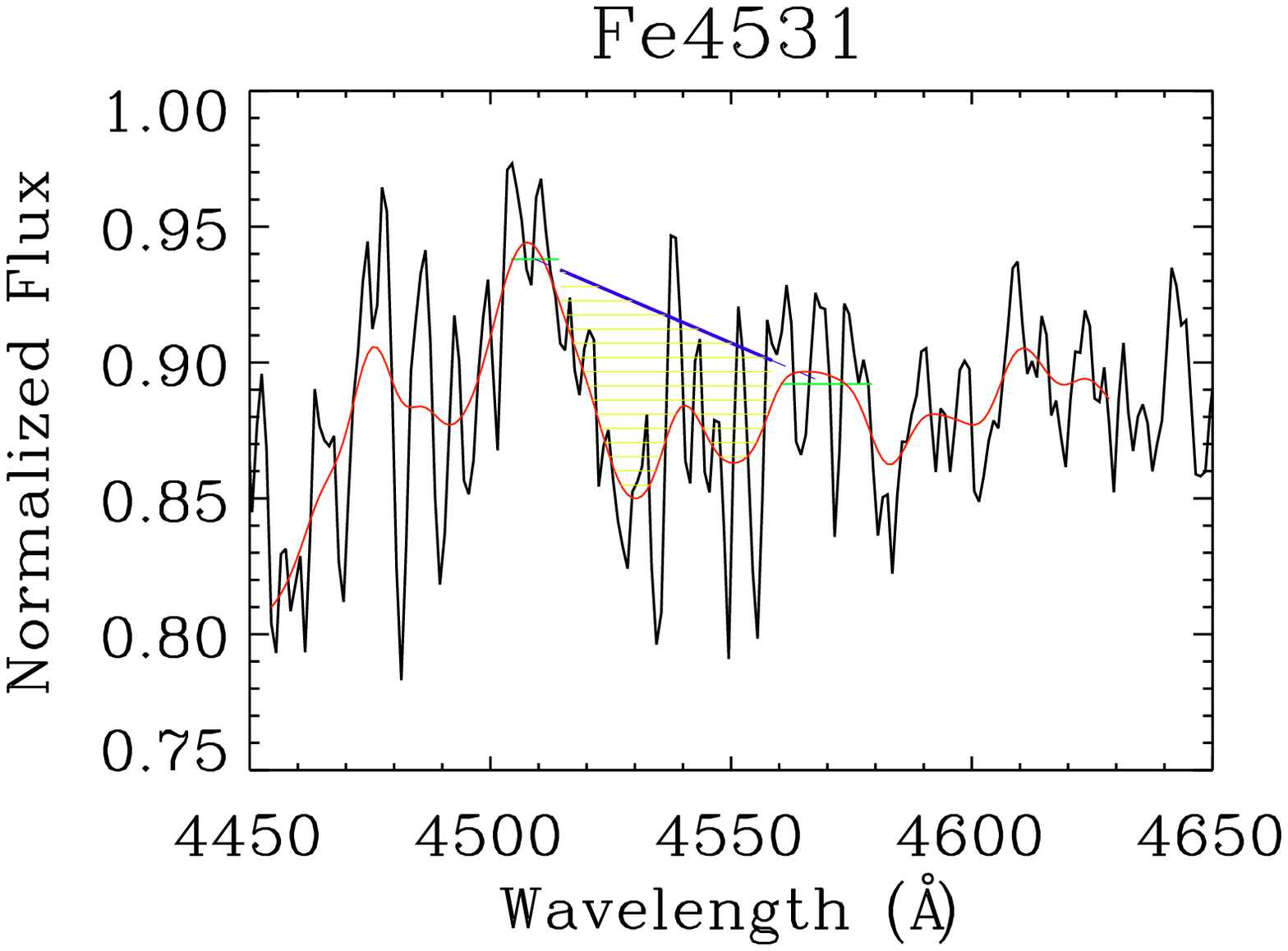}
\includegraphics[width=0.49\columnwidth]{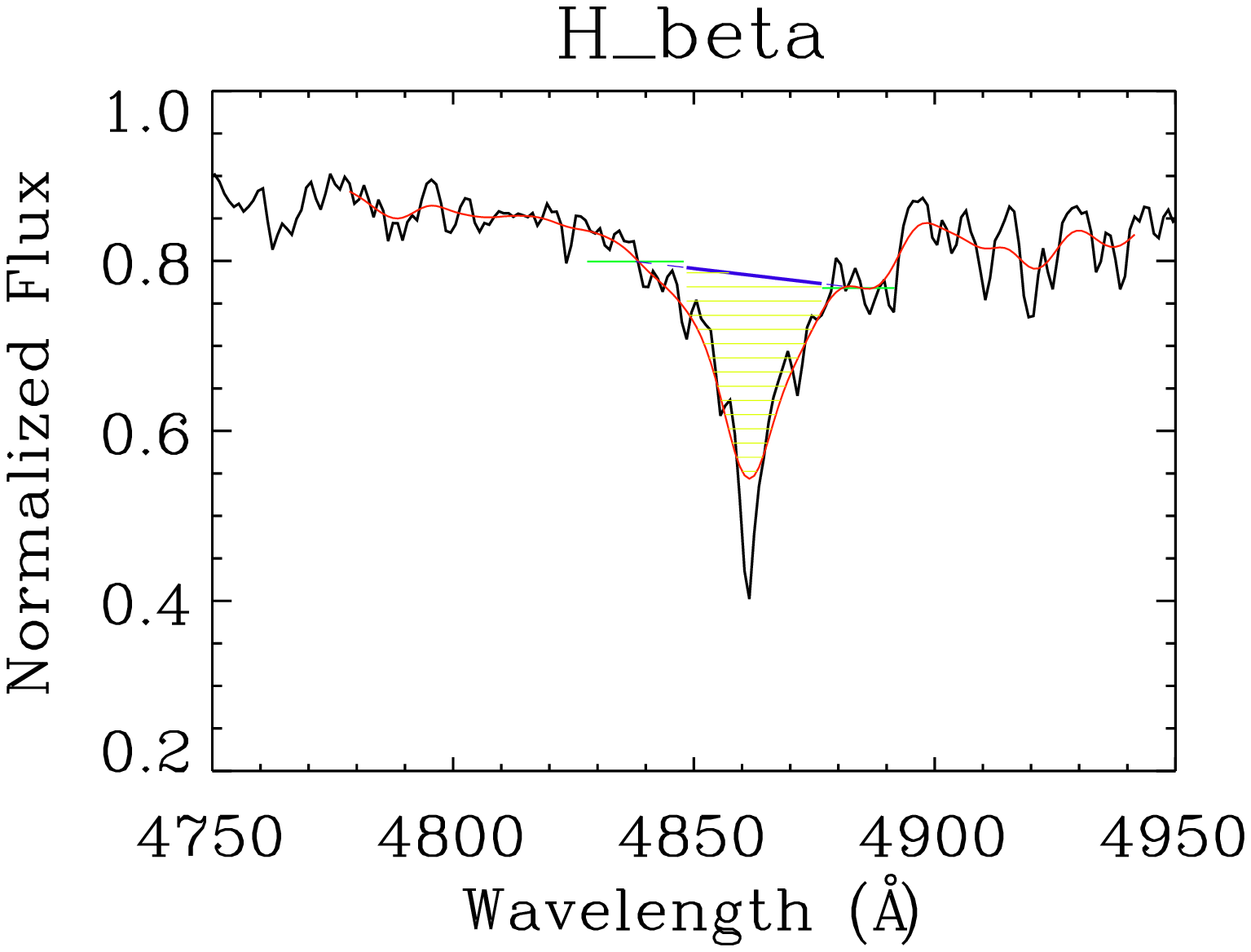}
\caption[Lick/IDS Index Defined Regions]{Six of the 12 Lick/IDS index regions highlighted on a BaSTI synthetic spectrum (black line, $Z = 0.01$ and age $=1$~Gyr SSP) utilized in fitting the age and metallicities of the cluster spectra.  The smoothed spectrum is overplotted in red.  The passband regions are within the solid blue lines (yellow lines filling the area used to measure $W_\lambda$), and the bounds of the pseudo continua are shown in green.  \label{LickIndices1}}
\end{center}
\end{figure}

\begin{figure}[htp]
\begin{center}
\includegraphics[width=0.49\columnwidth]{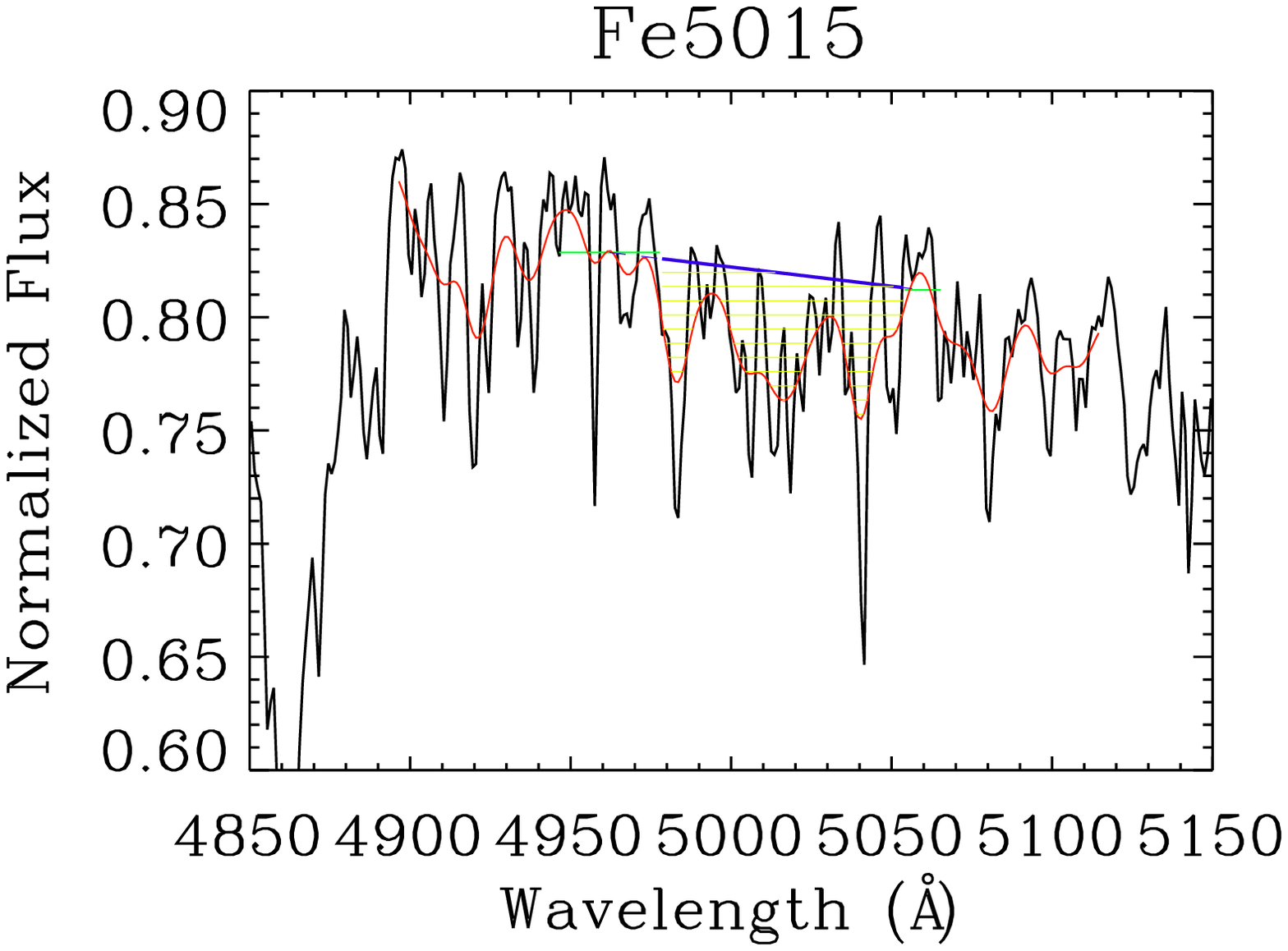}
\includegraphics[width=0.49\columnwidth]{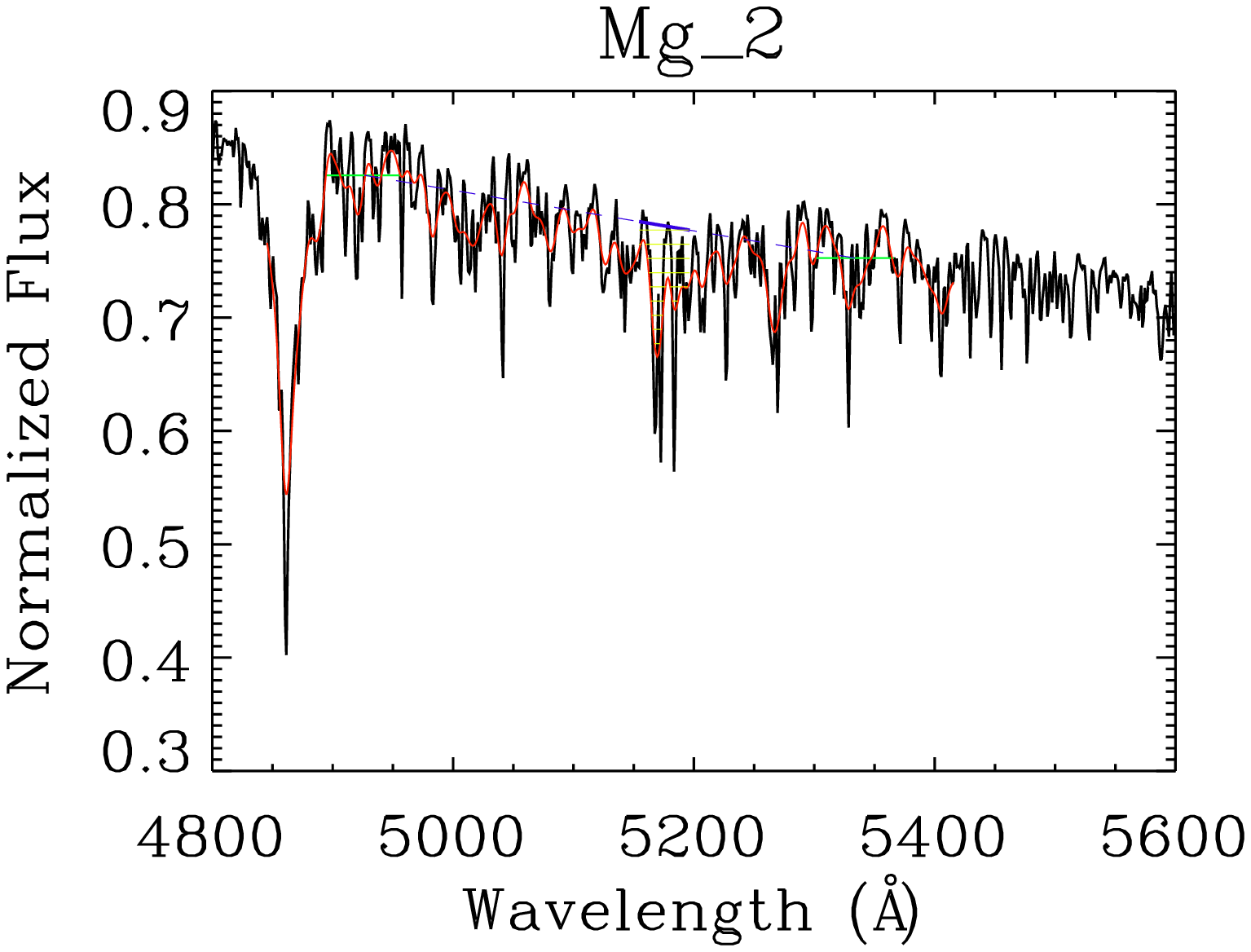}
\includegraphics[width=0.49\columnwidth]{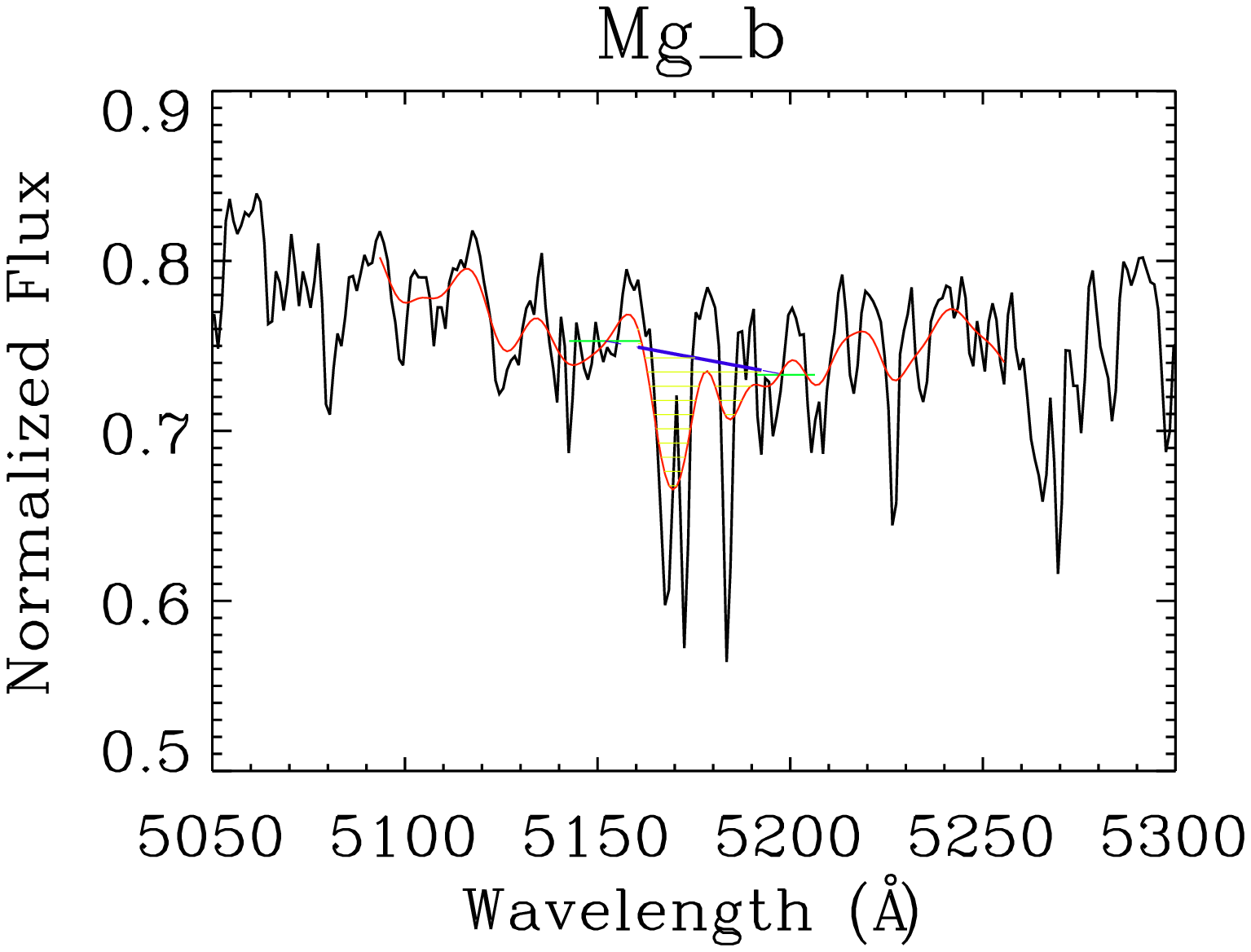}
\includegraphics[width=0.49\columnwidth]{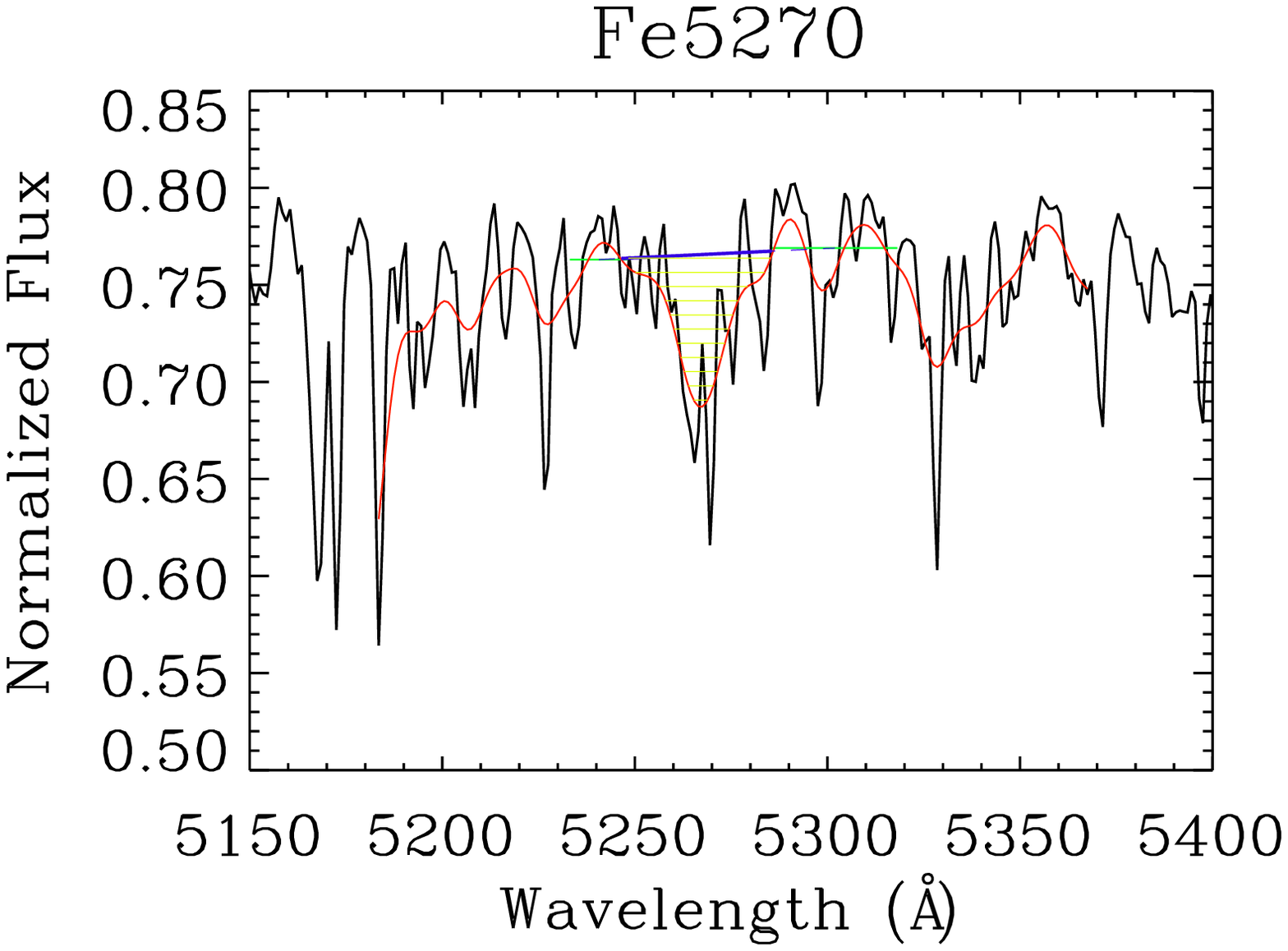}
\includegraphics[width=0.49\columnwidth]{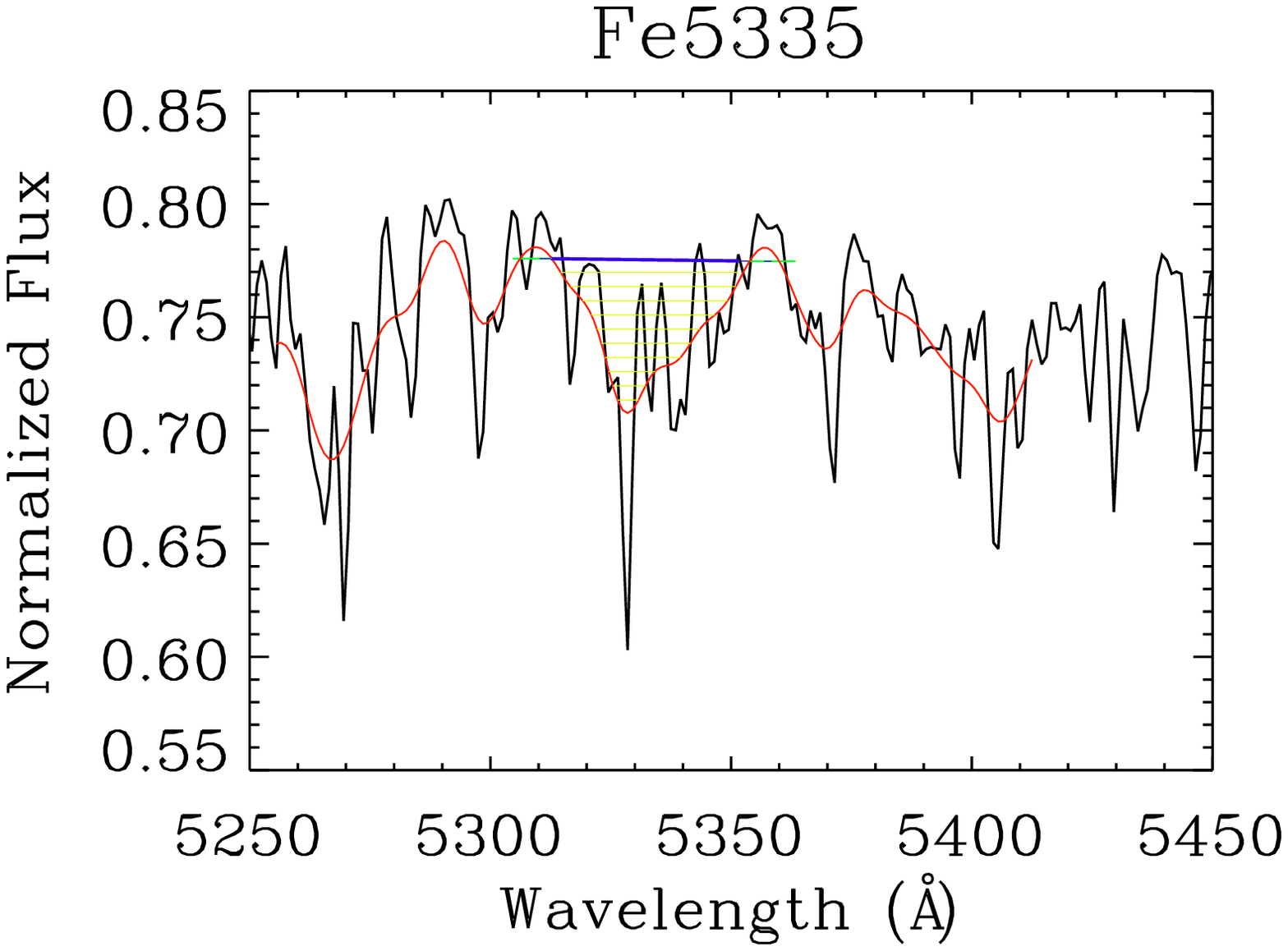}
\includegraphics[width=0.49\columnwidth]{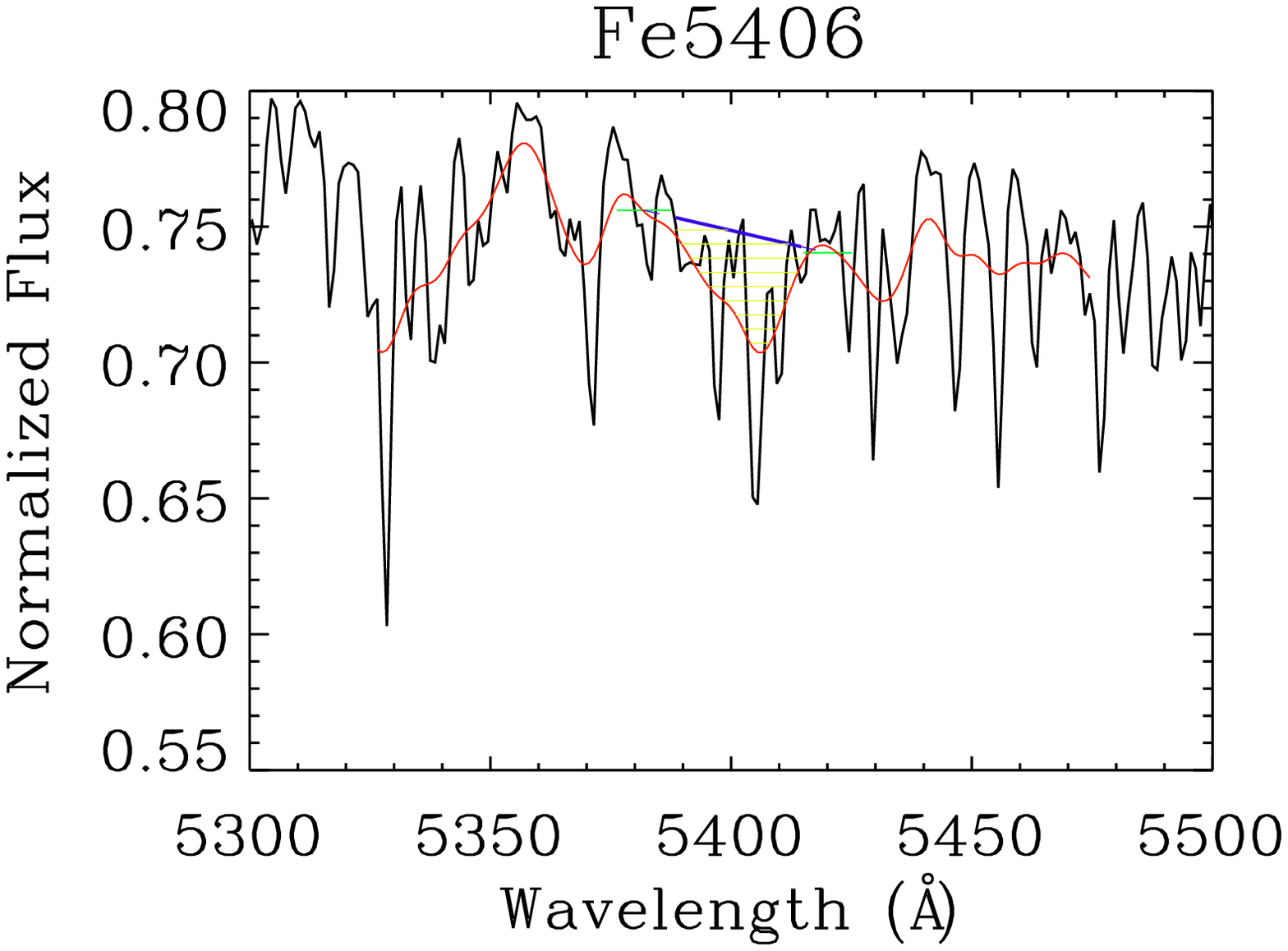}
\caption[Lick/IDS Index Defined Regions Continued]{The other six of the 12 Lick/IDS index regions highlighted on the same BaSTI synthetic spectrum with the same color coding as Fig.~\ref{LickIndices1}.  \label{LickIndices2}}
\end{center}
\end{figure}

Figures~\ref{LickIndices1} and \ref{LickIndices2} show the twelve Lick indices used here.  These indices are chosen because they are strong enough to be observable above the noise in most of the cluster spectra and are dependent on either age (H$\beta$, H$\gamma$A, H$\gamma$F, H$\delta$A, and H$\delta$F) or metallicity (Fe4531, Fe5015, Mg2, Mgb, Fe5270, Fe5335, and Fe5406).  Note that the $H\gamma$, $H\delta$, and Mg indices have different definitions with different widths of the passbands and/or pseudo continua.  Mg2 is considered a measurement of a molecular band, rather than an atomic band, with a magnitude index measurement, rather than $W_\lambda$ index in \AA, given by Mg2 $=-2.5~\text{log}[(\frac{1}{\lambda_2-\lambda_1})\int_{\lambda_1}^{\lambda_2}\frac{F_\lambda}{F_C}d\lambda]$.  Also, three of the metal indices are combined to form a single index, [MgFe]$^\prime = \sqrt{\text{Mgb}~(0.72~\text{Fe}5270 + 0.28~\text{Fe}5335)}$, known for being particularly insensitive to $\alpha$-element variations \citep{puz05}.

We choose BaSTI models with solar $\alpha$-element abundance\footnote{$\alpha$-elements (C, O, Ne, Mg, Si, S, Ar, and Ca) are created during the alpha process from He burning during nuclear fusion.} and \citet{rei75} red giant mass loss parameter $\eta = 0.4$ (to represent a blue horizontal branch) for comparison to the GC spectral indices.  We use $\eta = 0.2$ (to represent a red clump) for comparison to the YMC spectral indices.  The BaSTI library contains high resolution spectra (1.0~\AA$/$pixel), which we smooth to the resolution of the observed cluster spectra.  We utilize SSPs with ten different metallicities (Z $= 0.04$, 0.03, 0.0198 (solar), 0.01, 0.008, 0.004, 0.002, 0.001, 0.0006, and 0.0003 with [Fe/H] $= 0.40$, 0.26, 0.06, -0.25, -0.35, -0.66, -0.96, -1.27, -1.49, and -1.79) and $33-46$ ages per metallicity spanning from 50~Myr to 13.5~Gyr.

\begin{figure}[htp]
\begin{center}
\includegraphics[width=\columnwidth]{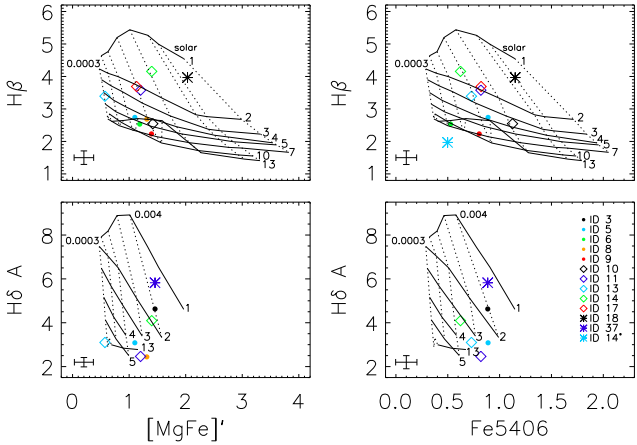}
\caption[Index-index Model Grids From BaSTI Synthetic Spectra for Old Ages]{Example index-index grids showing lines of constant ages (solid lines, from top to bottom: 1, 2, 3, 4, 5, 7, 10, and 13~Gyr) and constant Z (dotted lines, from left to right: 0.0003, 0.0006, 0.001, 0.002, 0.004, 0.008, 0.01, 0.0198 or [Fe/H]: -1.79, -1.49, -1.27, -0.96, -0.66, -0.35, -0.25, 0.06).  Index measurements for GCs are shown as colored solid circles, open diamonds, and asterisks with median error bars in the bottom left corner of each plot.  The asterisk denotes ID 14 on mask 2. \label{BaSTIoldGrids}}
\end{center}
\end{figure}

Figure~\ref{BaSTIoldGrids} shows some example index-index grids of constant ages and metallicities using two age sensitive Balmer indices and two metallicity sensitive indices for old ages and sub-solar metallicities.  The overlap of the 13~Gyr age line is caused by the effect of the mass loss parameter $\eta = 0.4$ which mimics clusters with a blue horizontal branch (HB), rather than setting $\eta = 0.2$ for clusters whose evolved red giant branch (RGB) stars form a red clump.  Bluer HBs generally correlate with lower metallicities; however, there is a ``second parameter" effect which causes some higher metallicity clusters to also have blue HBs \citep{ric97,swe98}.  Thus, despite the inconvenience of the overlapping 13~Gyr model, we choose to use the $\eta=0.4$ models for all GC measurements.  Note that the overlap of the 13~Gyr model is less pronounced at the lowest metallicities because blue HBs at these metallicities form extended tails down to faint magnitudes which contribute less total flux to the Balmer lines, despite having hotter temperatures.

\begin{figure}[htp]
\begin{center}
\includegraphics[width=\columnwidth]{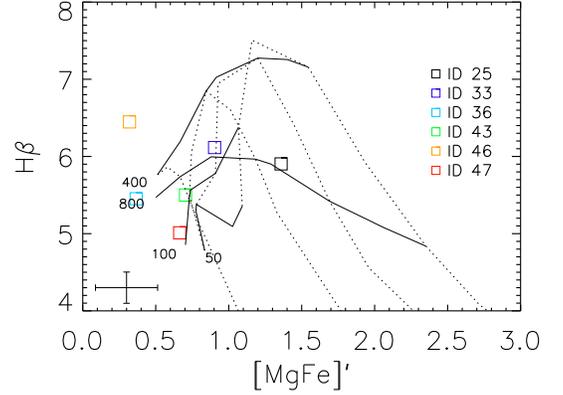}
\caption[Index-index Model Grids From BaSTI Synthetic Spectra for Young Ages]{Example index-index grid showing lines of constant ages (solid lines, labeled: 50, 100, 400, 800~Myr) and constant Z (crossing dotted lines: 0.002, 0.008, 0.0198 (solar), 0.04 or [Fe/H]: -0.96, -0.35, 0.06, 0.40).  Index measurements for YMCs are shown as colored open squares with median error bars in the bottom left corner. \label{BaSTIyoungGrids}}
\end{center}
\end{figure}

For determining the ages and metallicities of younger clusters we use Lick/IDS indices of SSPs with younger ages and include higher metallicity tracks.  We also use SSPs with $\eta = 0.2$ to represent RGB stars forming a red clump, which is the more likely evolutionary path for clusters formed at later times from higher metallicity gas.  Figure~\ref{BaSTIyoungGrids} shows an index-index plot and illustrates that at younger ages, the uniform grid configuration of the constant age and metallicity modeled indices breaks down.  It is still possible to fit age and metallicities to the YMC spectra; however, this requires a least $\chi^2$ fit to multiple indices at once \citep{tra07}.  This is a more robust method than plotting clusters on the index-index grids alone, and we employ it for both the YMCs and GCs.  

The example index measurements shown on Figures~\ref{BaSTIoldGrids} and \ref{BaSTIyoungGrids} illustrate that while small errors are calculated for each index measurement made with LICK\_EW, there are larger errors associated with the final cluster ages and metallicities due to differences between the resulting model fits for each index-index combination.  This further demonstrates the need for a robust fitting method to all of the indices simultanenously, as described below.

In order to estimate the age and metallicity of each cluster, we run a least $\chi^2$ minimization fit of each set of twelve cluster indices with a sufficiently high quality (based on individual visual inspection to remove lines with insufficient S/N) to the model BaSTI indices.  The $\chi^2$ is divided by the sum of the squares of the errors for each index output by the LICK\_EW code which are based on the S/N plane of each spectrum.

To add robustness to the fits, we run the $\chi^2$ minimization 5000 times for each cluster with artificial errors added to each index that follow a normal distribution with a width, $\sigma$, equal to the LICK\_EW error of the index.  We then take the final age and metallicities of the clusters as the mean value from the 5000 fits and then errors as the standard deviation of these 5000 runs. 

For the YMCs, we only fit to models with ages up to 2~Gyr, and for the GCs, we only fit models with ages down to 600~Myr and metallicities up to solar.  This appears to span the required range.  Also note that Fe5015 was of poor or questionable quality in all of the GC spectra, so it was not used.  We compare the age and metallicity results of the $\chi^2$ fits to the closest age and metallicity tracks on index-index plots, and find similar results which give us added confidence in the accuracy of the $\chi^2$ fits.  

Ages and metallicities of each cluster are shown in Table \ref{agemetalTable}.  Note that IDs 36 and 47 have strong Balmer absorption, and the best-fit ages are close to the youngest model in our grid, making it difficult to estimate the lower age uncertainty.  For these clusters, we therefore only provide an upper uncertainty on the age in Table \ref{agemetalTable}.

\vfill
\vfill

\vfill
\vfill

\vfill
\vfill

\vfill
\vfill

\input{table4}

\section{Results and Analysis}
\label{sec:Results}

\subsection{Velocity Distance Comparison}
\label{subsec:VelocityDistance}

\begin{figure}[htp]
\begin{center}
\includegraphics[width=\columnwidth]{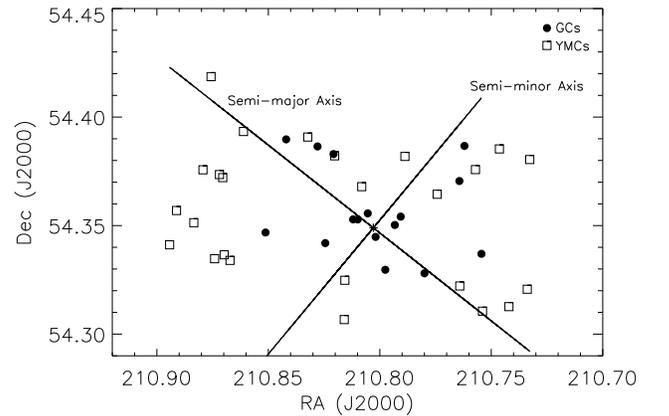}
\caption[Positions of the M101 YMCs and GCs with Spectra]{Positions of the YMCs (open squares) and GCs (solid circles) with respect to the semi-major and minor axes of M101 (labeled).  \label{specPositions}}
\end{center}
\end{figure}

\begin{figure}[htp]
\begin{center}
\includegraphics[width=\columnwidth]{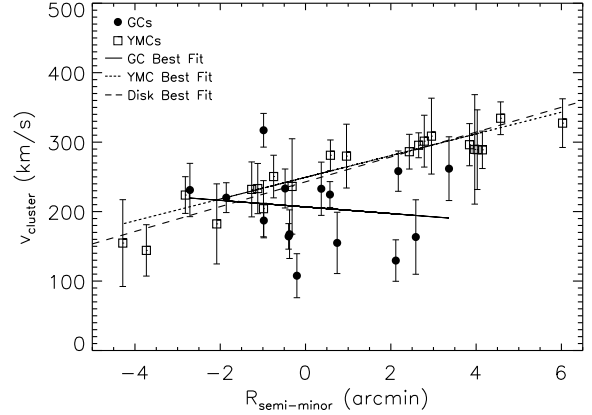}
\caption[$v_{\text{cluster}}$ vs. $R_\text{semi-minor}$ for the M101 YMCs and GCs with Spectra]{Velocities versus distance to the semi-minor axis of the YMCs (open squares) and GCs (solid circles).  The best fit lines to the GCs (solid line), YMCs (dotted line), and HI gas disk (dashed line, individual points not shown) are overplotted.  The YMCs have a similar although slightly shallower slope than the HI gas disk, while the GCs do not match the disk rotation (see Table~\ref{velradSlopesTable}). \label{velrad}}
\end{center}
\end{figure}

Figure~\ref{specPositions} shows the positions for each cluster with the semi-major and minor axes plotted (major axis position angle $\sim39^\circ$, \citet{bos81}).  From this plot we can obtain the perpendicular distance from each cluster to the semi-minor axis, $R_{\text{semi-minor}}$.  We then obtained estimates of the disk velocities, $v_{\text{disk}}$, for the RA and Dec of each cluster from a study of M101's HI gas by \citet{bos81} (see Table~\ref{kinTable}).  Figure~\ref{velrad} shows $v_{\text{cluster}}$ versus $R_{\text{semi-minor}}$.  Also shown are the linear best fits to the YMCs, GCs, and the HI gas disk velocities.  The slopes and errors of the best fits are found in Table~\ref{velradSlopesTable}.  Note that we exclude all clusters with $\sigma_{v_{\text{cluster}}}>100$~km/s (ID 42 on mask 1 and IDs 10*, 17*, 26*, and 28* on mask 2) from Fig~\ref{velrad} as well as the fits to the GCs and YMCs as these clusters have large uncertainties from noisy spectra and sometimes poor coverage, limiting the wavelength coverage available for the cross-correlation.

\vfill
\vfill

\vfill
\vfill

\input{table5}

\subsection{Age Velocity Comparison}
\label{subsec:AgeVelocity}

\begin{figure}[htp]
\begin{center}
\includegraphics[width=\columnwidth]{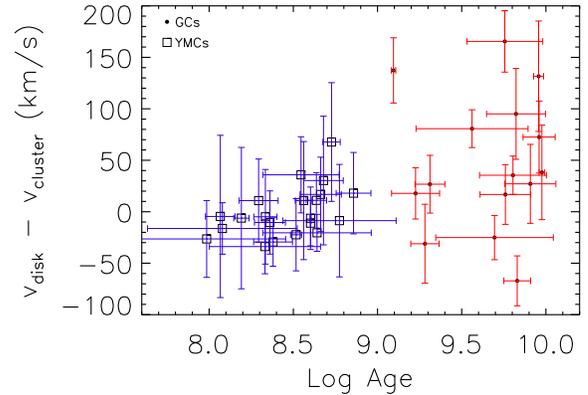}
\caption[$v_{\text{cluster}}-v_{\text{disk}}$ vs. log Age for M101 YMCs and GCs with Spectra]{The difference between the cluster and disk velocities versus ages of the YMCs (open squares) and GCs (solid circles).  \label{velage}}
\end{center}
\end{figure}

The difference between $v_{\text{disk}}$ and $v_{\text{cluster}}$ roughly corresponds to the distance of each cluster above or below the gas disk.  Thus, we show in Figure~\ref{velage} this velocity difference versus the age of the clusters.  The spread in velocity differences is greater for the GCs than the YMCs, and even within the YMC population, older YMCs have a broader spread than younger YMCs.  

The standard deviation of the velocity differences of the two populations are the best estimate of the velocity dispersions, $\sigma$, of the YMC and GC populations.  We find that $\sigma_{\text{YMC}} \approx 25$~km/s and $\sigma_{\text{GC}} \approx 66$~km/s.

\vfill
\vfill

\subsection{Rotational Velocity Calculation and Comparison}
\label{subsec:RotVelocity}

Because the inclination of M101 is not quite zero, i.e. not perfectly face-on, the rotational velocity can be calculated from the line-of-sight velocities for each star cluster associated with the disk and for the HI gas along the semi-major axis.  We determine the rotational velocities by $v_{\text{rot}} = (v_{\text{los}} - v_{\text{sys}})/\text{cos}~\theta~\text{sin}~i$ where $v_{\text{los}}$ is the line-of-sight velocity (either $v_{\text{cluster}}$ or $v_{\text{disk}}$ here), $v_{\text{sys}}$ is the systemic velocity of the galaxy ($\sim241$~km/s for M101), $\theta$ is the PA of each object with respect to the semi-major axis, and $i$ is the inclination of the galaxy ($\sim18^\circ$ for M101 \citep{bos81}).  In \S~\ref{subsec:VelocityDistance} and \ref{subsec:AgeVelocity}, we find that the GCs do not show evidence of association with the disk of M101, and therefore, the formula for $v_{\text{rot}}$ (dependent on sin~$i$) is not valid.  Thus, we do not include the GCs in our rotational velocity analysis.  

\begin{figure}[htp]
\begin{center}
\includegraphics[width=\columnwidth]{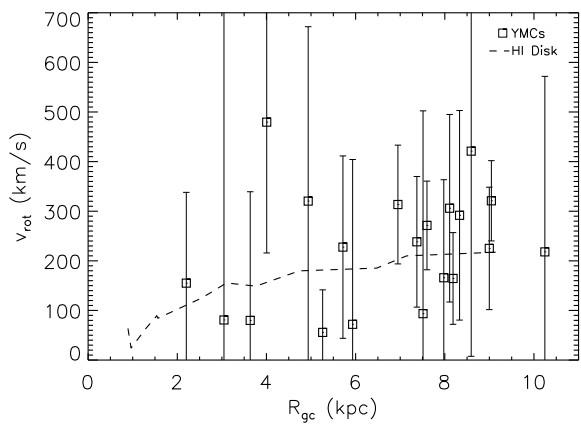}
\caption[$v_{\text{rot}}$ vs. $R_{\text{gc}}$ for the M101 YMCs and HI Gas]{The rotational velocities versus $R_{\text{gc}}$ for the YMCs (open squares) and HI gas (dashed line).  \label{vrot}}
\end{center}
\end{figure}

Table~\ref{vrotTable} shows the mean rotational velocities for the HI gas in the disk and the YMCs along with standard deviations as the errors.  We list the values for M101 populations in their entirety (excluding clusters with $\sigma_{v_{\text{cluster}}}>100$~km/s) as well as only for $R_{\text{gc}}>5$~kpc since $v_{\text{rot}}$ is the peak/plateau of a rotation curve, which is better described by the outer clusters.  We also list the values for the MW and M33 for comparison.  Figure~\ref{vrot} shows $v_{\text{rot}}$ as a function of distance, i.e. the rotation curves, for the YMCs and HI gas in the disk.

\vfill
\vfill

\input{table6}

\subsection{Age, Metal, and Spatial Distributions}
\label{subsec:SpatialMetals}

\begin{figure}[htp]
\begin{center}
\includegraphics[width=\columnwidth]{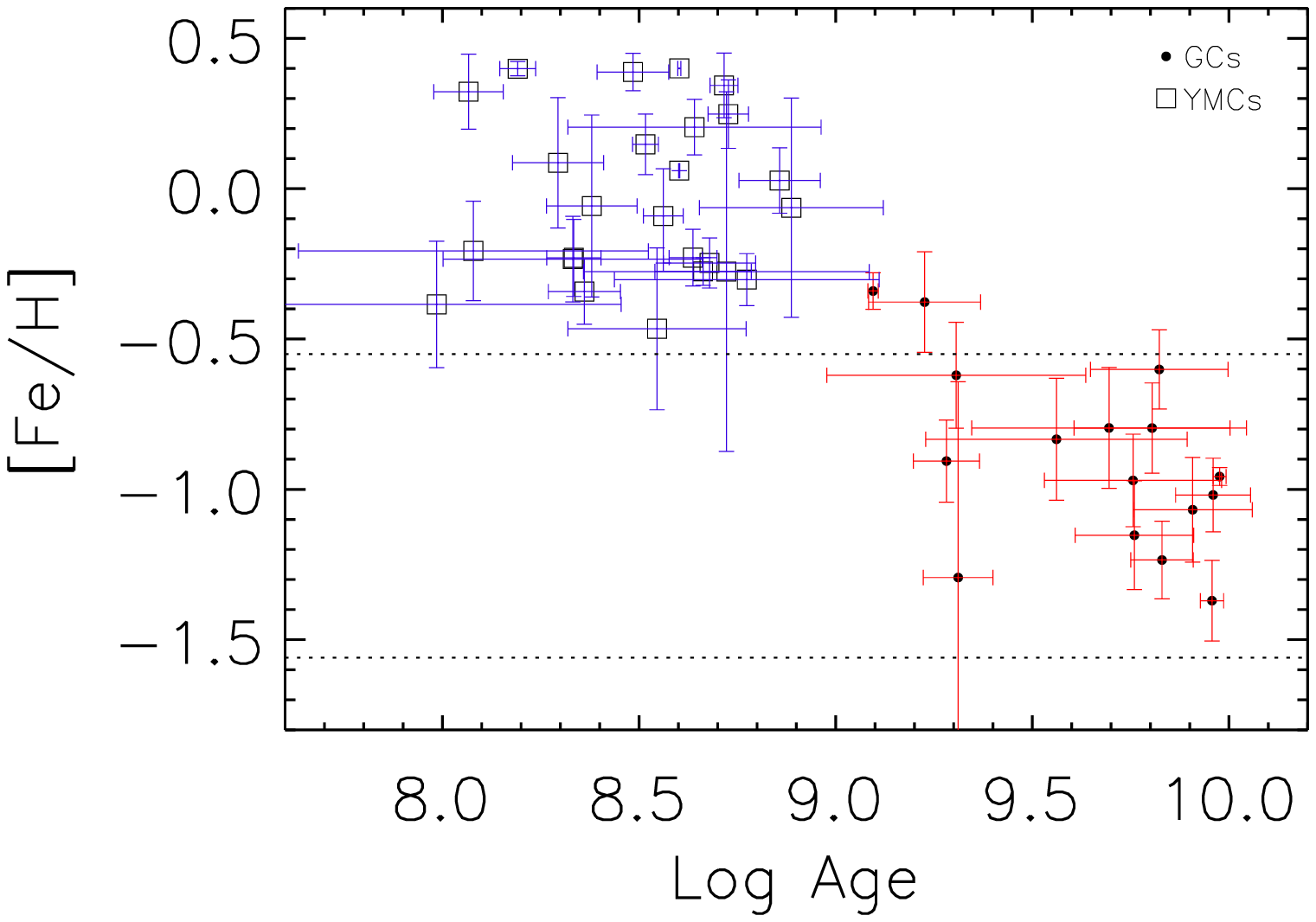}
\caption[{[Fe/H]} vs. log Age for M101 YMCs and GCs with Spectra]{[Fe/H] vs. log Age for all clusters.  Open squares with blue error bars represent YMCs, and solid circles with red error bars represent GCs.  The two dotted lines at [Fe/H] $= -0.55$ and $-1.56$, which correspond to the metal rich and poor peaks respectively of the MW GC system. \label{AgeFeFig}}
\end{center}
\end{figure}

Figure~\ref{AgeFeFig} shows the [Fe/H] versus the log of the ages for both GCs and YMCs with error bars over plotted.  The median age of the GCs is $5\pm2.9$~Gyr and $400\pm180$~Myr for the YMCs.  The ages of the GCs are much younger than typical GC ages in the MW ($\sim10$~Gyr).  Interestingly, our sample includes five GCs with young ages $\approx1-3$~Gyr (IDs 4, 14, 17, 18, and 28* on mask 2).  One of these clusters, ID 28* on mask 2, has a large error on its age due to having poor Balmer lines with no age lines included in the fit, and we cannot definitively conclude that this cluster is young.  

It is of note that ID 3 has starkly mismatched equivalent widths for the Balmer series of lines with H$\beta$ indicating a much younger age ($\sim800$~Myr) than H$\gamma$ or H$\delta$ indicate as well as very strong, sharp absorption lines at Fe5015 and $\lambda\approx4964$~\AA (see Fig.~\ref{GCspec1_mask1}).  In \S\ref{subsec:MeasureAgeMetals}, we excluded H$\beta$ from the age determination of ID 3, rather than the other three available Balmer Lick indices, since the cluster shows some observable G4300 absorption, which typically indicates an age of at least $\sim3$~Gyr.  Finally, IDs 11 and 37 have large errors on their ages which prevent us from definitively categorizing them as either young or old age GCs.  {\bf This leaves at least four GCs (IDs 4, 14, 17, and 18) with ages most likely within the young age range of $\sim1-3$~Gyr and eight GCs (IDs 5, 6, 8, 9, 10, 13, 15, and {\bf 14*} on mask 2) with definitively older ages of $5-9$~Gyr.}

\begin{figure}[htp]
\begin{center}
\includegraphics[width=\columnwidth]{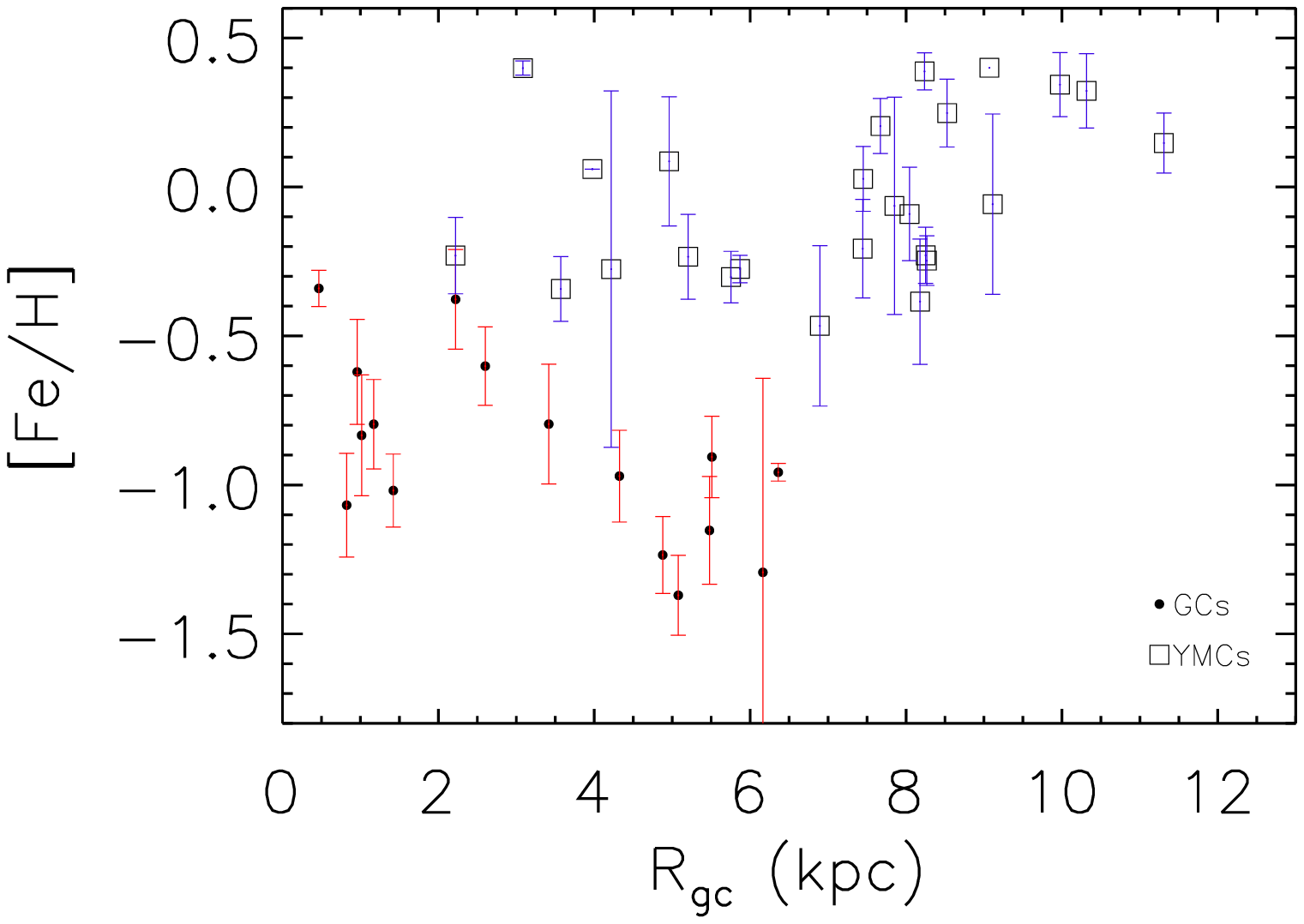}
\caption[{[Fe/H]} vs. $R_{\text{gc}}$ for M101 YMCs and GCs with Spectra]{[Fe/H] vs. $R_{\text{gc}}$ for all clusters.  Open squares with blue error bars represent YMCs, and solid circles with red error bars represent GCs.  \label{DistFeFig}}
\end{center}
\end{figure}

The median [Fe/H] for the GCs is $-0.91\pm0.30$ and $-0.06\pm0.27$ for the YMCs.  Figure~\ref{DistFeFig} shows the [Fe/H] versus the galactocentric distance of each cluster from the center of M101 \citep{eva10} as projected onto the plane of the sky ($R_{\text{gc}}$).  Note that the GCs appear more centrally concentrated, while the YMCs are spread to further distances, limited by the extent of the two GMOS masks for the observations.  We discuss the possible conclusions drawn from this metallicity distribution in \S\ref{subsec:AgeMetalYMCsGCs}.

\vfill
\vfill

\vfill
\vfill

\section{Discussion}
\label{sec:Discussion}

\subsection{The Structure of M101}
\label{subsec:Structure}

We find that the YMCs follow the HI gas in the disk (see Figure~\ref{velrad} and Table~\ref{velradSlopesTable}).
The $v_{\text{rot}}/\sigma$ ratio (see Table~\ref{vrotTable}) also supports disk-like rotation.
The YMCs do however, show some scatter about the HI disk fitted $v_{\text{cluster}}$-$R_{\text{semi-minor}}$ relation, a lower $v_{\text{rot}}/\sigma$ ratio than the HI gas, and have a $v_{\text{rot}}/\sigma$ value that is comparable to that of the old ($\geq1$~Gyr) open clusters in the MW rather than younger disk clusters.  Thus, the YMCs have similar rotation to the HI disk, but a larger velocity dispersion.  Interestingly, $\sigma_{\text{YMC}}$ is similar to $\sigma_{\text{pseudobulge}}$ determined by \citet{kor10} ($\sim 25$~km/s compared to $27\pm4$~km/s), which supports the possibility that the central portion of M101 is more of an inner disk than a bulge.

The line-of-sight velocities for the GCs do not follow the HI gas disk (despite having lower $v_{\text{cluster}}$ errors than the YMCs).  
While the velocities are more similar to those expected for a halo than for a thin disk, the velocity dispersion of the GCs in M101 is more similar to the values found for MW GCs associated with the thick disk and/or bulge.  We cannot rule out the possibility that the GCs sampled here form a pseudobulge or thick disk population; however, the effective radius determined from light profile fitting of the pseudobulge ($r_{e}=400_{-300}^{+800}$~pc, \citet{fis10}) is small enough to call into question the idea that these GCs (some at distances out to more than 6~kpc) are associated with the bulge alone.  There are most likely some halo GCs (also supported by the more complete photometric GC catalog discussed in \citet{sim15}).  Their existence potentially conflicts with the results of \citet{van14} who found M101 contains little to no halo component.

Figure~\ref{velage} shows that the clusters in M101 have a continuous increase in the dispersion of the velocity residuals (with respect to the disk velocities) with age.  This is similar to the trend seen for M33 clusters \citep{cha02}, and indicates that systems of older clusters have undergone a source of ``heating" whether by perturbations from passing giant molecular clouds (a secular process) or mergers/accretion.  The larger velocity dispersion of the YMCs compared to the HI disk also supports this conclusion.  

Specific external and internal sources of heating/disturbance of the M101 disk are discussed in \citet{wal97}.  They found from comparing far ultra-violet imaging of M101 to simulations that M101 has undergone interactions with companion galaxies, possibly NGC 5477 and NGC 5474, within the past $10^8-10^9$~Gyr forming a tidal tail and supergiant HII region, NGC 5471.  This time scale matches well with our YMC ages and possibly with three of the four GCs with ages very close to 1~Gyr.  \citet{wal97} further proposed that NGC 5471 and other massive HII regions have caused crooked spiral arms and linear arm segments throughout the disk of M101, making internal sources of disk heating possible as well.

\subsection{Ages and Metallicities of the YMCs and GCs}
\label{subsec:AgeMetalYMCsGCs}

From Figure~\ref{AgeFeFig}, we see that the ages for the sample of M101 star clusters is almost continuous, most surprisingly even for ages greater than 1~Gyr.  We find that there are both ``young'' and ``old'' GCs.
Figure~\ref{AgeFeFig} also shows that the metallicities of the YMCs and GCs roughly separate into metal-rich and metal-poor populations (as expected if the YMCs formed more recently out of metal enriched gas in the disk of M101 and the GCs in the spheroidal components formed earlier out of more metal-poor gas).  However, the transition between the young to old clusters follows an almost continuous trend of decreasing metallicity.  This hints at more or less continuous cluster formation, possibly aided by mergers/interactions, in M101's past.  

Seven of the GCs have metallicities completely above (including error bars) the lower limit of the radial abundance gradient observed for MW bulge GCs ([Fe/H] $=-1.0$, \citet{min95_2}).  While the median of all of the GC metallicities is [Fe/H] $= -0.91 \pm 0.30$, dividing our sample along [Fe/H] $=-1.0$ (seven clusters completely above, including error bars, and nine clusters at or below [Fe/H] $=-1.0$) gives a metal-poor GC mean [Fe/H] $=-1.1 \pm 0.2$ and metal-rich mean [Fe/H] $=-0.6 \pm 0.2$.  While this division is artificially imposed, it suggests that these GCs may be sampling both a bulge/thick disk population of GCs as well as an older, more metal poor halo population.  

From Figure~\ref{DistFeFig} we see that the GCs studied here are entirely within the inner portions of M101 with $R_{\text{gc}}<8$~kpc.  Three of the seven GCs with [Fe/H] $> -1.0$ (IDs 10 and 14 on mask 1, and ID 28* on mask 2) and two of the nine GCs with [Fe/H] $< -1.0$ (IDs 8 and 11) are at $R_{\text{gc}}<1.2$~kpc, which is the upper limit of the effective radius of the bulge of M101, $r_{e}=400_{-300}^{+800}$~pc \citep{fis10}.  While the presence of clusters with a spread in metallicities (including metal-rich clusters) could support a bulge formed through mergers rather than by secular processes, our sample size is not sufficiently large enough to rule out a secularly formed pseudo-bulge, which is the type of bulge expected for M101 given its morphology.

The nine metal-poor GCs have a mean [Fe/H] greater than observed for the MW GCs ([Fe/H] $\approx-1.5$).  These GCs are also overall younger (mean age $\sim6 \pm 3$~Gyr) than the very ancient MW GCs ($\sim13$~Gyr).  This combination of elevated metallicities and an age spread encompassing younger ages may indicate a rich history of mergers and accretion in the halo of M101, which would make  it more difficult to differentiate between populations of GCs in a thick disk versus in the inner halo.

\section{Conclusions}
\label{sec:Conclusions}

We find that the massive star clusters in the spiral galaxy M101 have a fairly continuous spread of ages and metallicities.  The YMC and GC populations separate into younger and more metal-rich (ages of hundreds of Myr and median [Fe/H] $\approx-0.1$) versus older and more metal-poor populations (ages from $1-12$~Gyr and median [Fe/H] $\approx-0.9$) as expected for a typical spiral galaxy structure of disk and spheroidal components.  However, the transition between the two categories of clusters is not as sharp as in the Milky Way, and there are at least four GCs which appear to be quite young, with best fit ages $\sim1-3$~Gyr.
The best fit ages of the other GCs range from $\sim5$ to 10~Gyr, and reach the very old ages typical of GCs in our own Galaxy.

Our overall conclusion is that M101 had a rich, possibly continuous history of cluster and star formation.  The metallicities and spatial distribution of the GCs indicate that cluster formation is most likely driven by mergers/accretion.

We find the kinematics of the YMC and GC populations differ, with the YMCs following the characteristics of the HI gas in the disk and the GCs exhibiting line-of-sight and rotational velocities and velocity dispersion similar to either a bulge/thick disk or halo.  The smooth increase in the difference between the cluster velocities and their local disk velocities with age indicates that M101 may have undergone heating of its disk over time or a fairly continuous merger/accretion history.

We acknowledge support from the NSF through CAREER award 0847467, and we thank Genevieve J. Graves for making the code LICK\_EW available at \url{http://w.astro.berkeley.edu/~graves/ez_ages.html}.  Based on observations obtained at the Gemini Observatory (part of Gemini program GN-2008A-Q-55, processed using the Gemini IRAF package), which is operated by the Association of Universities for Research in Astronomy, Inc., under a cooperative agreement with the NSF on behalf of the Gemini partnership: the National Science Foundation (United States), the National Research Council (Canada), CONICYT (Chile), Ministerio de Ciencia, Tecnolog\'{i}a e Innovaci\'{o}n Productiva (Argentina), and Minist\'{e}rio da Ci\^{e}ncia, Tecnologia e Inova\c{c}\~{a}o (Brazil).

\bibliography{ArticleFile_submitArXiv.bib}
\bibliographystyle{apj}

\end{document}

%% file: table1.tex
\begin{deluxetable*}{lrrrrrrr}
\tablecolumns{8}
\tablewidth{0pc}
\tablecaption{Basic Properties for Our Spectroscopic Sample of Star Clusters in M101}
\tablehead{\colhead{ID} & \colhead{$\alpha_{2000}$~(hms)} & \colhead{$\delta_{2000}$~(\arcdeg\arcmin\arcsec)} & \colhead{$M_V$} & \colhead{$B-V$} & \colhead{$V-I$} & \colhead{$B-I$} & \colhead{$r_{\text{eff}}$~(pc)}}
\startdata
Mask 1 & & & & & & & \\
\hline
3\tablenotemark{a} & 210 50 31.24 & 54 23 23.02 & -8.01 & 0.67 & 1.03 & 1.70 & 2.50\\
4 & 210 45 43.27 & 54 23 12.14 & -8.00 & 0.75 & 1.09 & 1.83 & 1.70\\
5 & 210 49 14.66 & 54 22 58.62 & -9.06 & 0.65 & 0.96 & 1.61 & 1.68\\
6 & 210 45 51.48 & 54 22 13.94 & -7.86 & 0.70 & 1.00 & 1.70 & 2.34\\
8 & 210 48 19.44 & 54 21 20.22 & -8.62 & 0.61 & 0.97 & 1.58 & 3.81\\
9 & 210 47 26.27 & 54 21 14.88 & -8.48 & 0.79 & 1.08 & 1.87 & 1.26\\
10 & 210 48 43.13 & 54 21 10.32 & -8.52 & 0.82 & 1.19 & 2.01 & 1.91\\
11 & 210 47 35.81 & 54 21 1.02 & -9.18 & 0.99 &	1.30 & 2.29 & 1.77\\
13 & 210 51 4.50 & 54 20 48.49 & -8.97 & 0.66 & 1.01 & 1.67 & 3.12\\
14 & 210 48 6.77 & 54 20 41.27 & -7.95 & 0.65 & 1.00 & 1.66 & 1.54\\
15\tablenotemark{a} & 210 49 28.09 & 54 20 30.94 &  -8.27 & 0.72 & 1.06 & 1.78 & 2.50\\
17  & 210 45 15.88 & 54 20 13.19 & -8.28 & 0.63 & 0.99 & 1.62 & 2.16\\
18  & 210 47 51.18 & 54 19 46.67 & -8.03 & 0.66 & 1.05 & 1.71 & 2.25\\
23\tablenotemark{a} & 210 49 56.32 & 54 23 26.87 & -8.24 & 0.12 & 0.61 & 0.73 & 1.17\\
25 & 210 44 46.86 & 54 23 7.14 & -7.47 & 0.30 & 0.63 & 0.93 & 1.70\\
26 & 210 47 19.43 & 54 22 54.72 & -7.67 & 0.26 & 0.54 & 0.79 & 3.74\\
27 & 210 43 57.83 & 54 22 49.59 & -7.11 & 0.21 & 0.53 & 0.74 & 5.12\\
29\tablenotemark{a} & 210 52 18.95 & 54 22 24.63 & -7.57 & 0.35 & 0.70 & 1.05 & 8.70\\
30\tablenotemark{a} & 210 52 13.62 & 54 22 19.63 & -7.28 & 0.43 & 0.82 & 1.25 & 7.99\\
31\tablenotemark{a} & 210 48 29.16 & 54 22 4.57 & -8.32 & 0.12 & 0.40 & 0.51 & 3.24\\
33 & 210 46 27.44 & 54 21 51.99 & -8.74 & 0.14 & 0.50 & 0.64 & 2.82\\ 
36\tablenotemark{a} & 210 52 1.96 & 54 20 2.28 & -9.40 & 0.12 & 0.46 & 0.58 & 2.96\\
37 & 210 46 47.82 & 54 19 40.88 & -7.48 & 1.37 & 2.04 & 3.40 & 1.77\\
38 & 210 48 56.41 & 54 19 29.34 & -7.34 & 0.03 & 0.36 & 0.39 & 3.28\\
39 & 210 45 50.87 & 54 19 19.84 & -7.88 & 0.20 & 0.68 & 0.87 & 4.66\\
40 & 210 44 1.90 & 54 19 14.48 & -7.38 & 0.17 & 0.68 & 0.85 & 2.04\\
42 & 210 44 31.42 & 54 18 45.59 & -7.40 & 0.03 & 0.39 & 0.42 & 1.81\\
43 & 210 45 14.11 & 54 18 38.04 & -8.29 & -0.07 & 0.23 & 0.16 & 3.88\\
44 & 210 48 57.42 & 54 18 24.12 & -7.56 & 0.24 & 0.55 & 0.79 & 5.37\\
46 & 210 45 25.56 & 54 22 32.89 & -7.84 & 0.25 & 0.57 & 0.82 & 5.95\\
47 & 210 52 26.76 & 54 20 5.03 & -8.82 & 0.15 & 0.62 & 0.77 & 3.17 \\
\hline
Mask 2 & & & & & & & \\
\hline
5* & 210 52 32.52 & 54 25 7.05 & -7.26 & 0.14 & 0.39 & 0.54 & 5.01 \\
10* & 210 51 40.14 & 54 23 36.06 & -6.82 & 0.17 & 0.61 & 0.78 & 3.77\\
14* & 210 49 40.58 & 54 23 11.27 & -7.51 & 0.90 & 1.32 & 2.22 & 4.18\\
17* & 210 49 12.94 & 54 22 55.66 & -7.65 & 0.09 & 0.49 & 0.58 & 3.33\\
18* & 210 52 45.55 & 54 22 32.52 & -7.76 & 0.17 & 0.50 & 0.67 & 6.70\\
26* & 210 53 28.32 & 54 21 24.96 & -7.67 & 0.14 & 0.39 & 0.54 & 3.17\\
28* & 210 48 35.68 & 54 21 10.22 & -7.79 & 0.61 & 1.03 & 1.64 & 2.92\\
29* & 210 53 0.31 & 54 21 4.77  & -8.53 & 0.53 & 1.02 & 1.55 & 0.05\\
34* & 210 53 39.41 & 54 20 28.11 & -7.93 & 0.17 & 0.58 & 0.76 & 4.04\\  
36* & 210 52 11.32 & 54 20 11.58 & -8.02 & 0.15 & 0.63 & 0.78 & 4.43\\
\enddata
\tablenotetext{a}{Also on Mask 2.}
\label{photTable}
\end{deluxetable*}

%% file: table2.tex
\begin{deluxetable}{lccccl}
\tablecolumns{6}
\tablewidth{0pc}
\tablecaption{Basic Spectral Properties of Our Cluster Sample}
\tablehead{\colhead{ID} & \colhead{$\lambda_{\rm{min}}$-$\lambda_{\rm{max}}$} & \colhead{Count\tablenotemark{a}} & \colhead{S/N\tablenotemark{b}} & \colhead{$W_{\text{H}\beta}$ (\AA)} & \colhead{Type}}
\startdata
Mask 1 & & & & & \\
\hline
3\tablenotemark{c} & 3190-6090 & 1056 & 33 & 8.1 & GC \\
4 & 4050-7070 & 818 & 29 & 4.2 & GC \\
5 & 3440-6340 & 4345 & 66 & 3.1 & GC \\
6 & 4065-6960 & 991 & 32 & 3.8 & GC \\
8 & 3605-6550 & 2252 & 48 & 3.6 & GC \\
9 & 3605-6550 & 1968 & 44 & 3.4 & GC \\
10 & 3540-6440 & 3073 & 55 & 4.0 & GC \\
11 & 3750-6645 & 5154 & 72 & 3.5 & GC \\
13 & 3105-5990 & 3141 & 56 & 3.3 & GC \\
14 & 3650-6570 & 1136 & 34 & 5.4 & GC \\
15\tablenotemark{c} & 3400-6330 & 2367 & 49 & 0.4 & GC \\
17 & 4185-7075 & 1399 & 37 & 5.1 & GC \\
18 & 3660-6620 & 1127 & 33 & 4.0 & GC \\
23\tablenotemark{c} & 3300-6200 & 1616 & 40 &  5.9 & YMC \\
25 & 4250-7155 & 773 & 28 & 10.0 & YMC \\
26 & 3770-6725 & 1156 & 34 & 13.7 & YMC \\
27 & 4410-7320 &  456 & 21 & 11.5 & YMC \\
29\tablenotemark{c} & 3000-5730 & 2333 & 48 & 10.9 & YMC \\
30\tablenotemark{c} & 3000-5760 & 740 & 27 & 14.3 & YMC \\
31\tablenotemark{c} & 3570-6590 & 2162 & 47 & 10.2 & YMC \\
33 & 3910-6865 & 3192 & 57 &  9.3 & YMC \\
36\tablenotemark{c} & 2970-5780 & 6304 & 79 & 8.5 & YMC \\
37 & 3880-6820 & 890 & 30 & 1.1 & GC \\
38 & 3500-6390 & 666 & 26 & 8.7 & YMC \\
39 & 4060-6980 & 1632 & 40 & 12.1 & YMC \\ 
40 & 4365-7305 & 808 & 29 & 11.4 & YMC \\
42 & 4310-7210 & 497 & 22 & 5.3 & YMC \\
43 & 4190-7070 & 2411 & 49 & 8.1 & YMC \\
44 & 3490-6375 & 1739 & 42 & 9.6 & YMC \\
46 & 4140-7075 & 1584 & 40 & 9.6 & YMC \\ 
47 & 3000-5680 & 6464 & 80 & 6.9 & YMC \\
\hline
Mask 2 & & & & & \\
\hline
5* & 3730-6615 & 972 & 31 & 9.9 & YMC \\
10* & 3900-6780 & 538 & 23 & 8.9 & YMC \\
14* & 4260-7150 & 921 & 30 & 3.6 & GC \\
17* & 4340-7230 & 1840 & 43 & 8.7 & YMC \\
18* & 3690-6585 & 2256 & 48 & 9.7 & YMC \\ 
26* & 3555-6450 & 1208 & 35 & 14.3 & YMC \\
28* & 4435-7350 & 1396 & 37 & 5.2 & GC \\
29* & 3620-6540 & 1384 & 37 & 7.4 & YMC \\
34* & 3510-6460 & 1832 & 43 & 11.9 & YMC \\
36* & 3790-6690 & 2043 & 45 & 8.5 & YMC \\ 
\enddata
\tablenotetext{a}{Non-calibrated flux count at 5000 \AA.}
\tablenotetext{b}{S/N estimated by $\frac{N}{\sqrt[]{N}}$ where $N$ is the flux of the continuum at 5000 \AA.}
\tablenotetext{c}{Also on mask 2.}
\label{specTable}
\end{deluxetable}

%% file: table3.tex
\begin{deluxetable}{lcccc}
\tablecolumns{5}
\tablewidth{0pc}
\tablecaption{Kinematics of Our Cluster Sample}
\tablehead{\colhead{ID} & \colhead{$v_{\text{cluster}}$~(km/s)} & \colhead{$\sigma_{v_{\text{cluster}}}$~(km/s)} & \colhead{$v_{\text{disk}}$~(km/s)}}
\startdata
Mask 1 & & \\
\hline
3\tablenotemark{a} & 262 & 46 & 300 \\
4 & 233 & 28 & 260 \\
5 & 130 & 30 & 295 \\
6 & 317 & 24 & 250 \\
8 & 233 & 38 & 260 \\
9 & 167 & 35 & 240 \\
10 & 225 & 19 & 260\\
11 & 164 & 18 & 245 \\
13 & 258 & 29 & 275 \\
14 & 108 & 32 & 245 \\
15\tablenotemark{a} & 155 & 44 & 250\\
17 & 231 & 38 & 200 \\
18 & 187 & 25 & 205 \\
23\tablenotemark{a} & 309 & 55 & 300 \\
25 & 232 & 40 & 250 \\
26 & 281 & 22 & 270 \\
27 & 182 & 58 & 250 \\
29\tablenotemark{a} & 289 & 27 & 300 \\
30\tablenotemark{a} & 289 & 57 & 300 \\
31\tablenotemark{a} & 280 & 46 & 275 \\ 
33 & 250 & 31 & 240 \\
36\tablenotemark{a} & 286 & 25 & 270 \\
37 & 220 & 21 & 195 \\
38 & 236 & 69 & 230 \\
39 & 224 & 26 & 190 \\
40 & 155 & 63 & 185 \\
42 & 172 & 158 & 180 \\
43 & 144 & 37 & 180 \\
44 & 204 & 40 & 215 \\
46 & 233 & 36 & 250 \\
47 & 301 & 37 & 275\\
\hline
Mask 2 & & \\
\hline
5* & 327 & 35 & 305 \\
10* & 227 & 108 & 305 \\
14* & 163 & 54 & 295 \\
17* & 408 & 119 & 295 \\
18* & 334 & 24 & 305 \\
26* & 263 & 293 & 295 \\
28* & 225 & 119 & 265 \\
29* & 296 & 30 & 290 \\
34* & 290 & 79 & 285 \\
36* & 295 & 18 & 275 \\
\enddata
\tablenotetext{a}{Also on mask 2.}
\label{kinTable}
\end{deluxetable}

%% file: table4.tex
\begin{deluxetable}{lccccc}
\tablecolumns{6}
\tablewidth{0pc}
\tablecaption{Ages and Metallicities of Our Cluster Sample}
\tablehead{\colhead{ID} & \colhead{Age~(Myr)} & \colhead{$\sigma_{age}$~(Myr)} & \colhead{Z} & \colhead{[Fe/H]} & \colhead{$\sigma_{[Fe/H]}$}}
\startdata
Mask 1 & & & & & \\
\hline
3\tablenotemark{a} & 9500 & 360 & 0.0020 & -0.96 & 0.03 \\
4 & 2000 & 420 & 0.0028 & -1.29 & 0.65 \\
5 & 5700 & 2960 & 0.0021 & -0.97 & 0.15 \\
6 & 6800 & 1240 & 0.0011 & -1.24 & 0.13 \\
8 & 8100 & 2820 & 0.0017 & -1.07 & 0.17 \\
9 & 9100 & 2000 & 0.0018 & -1.02 & 0.12 \\
10 & 6400 & 2910 & 0.0031 & -0.80 & 0.15 \\
11 & 3600 & 2790 & 0.0030 & -0.83 & 0.20 \\
13 & 5700 & 1990 & 0.0014 & -1.15 & 0.18 \\
14 & 1240 & 40 & 0.0083 & -0.34 & 0.06 \\
15\tablenotemark{a} & 6600 & 2680 & 0.0048 & -0.60 & 0.13 \\
17 & 1910 & 370 & 0.0024 & -0.91 & 0.14 \\
18 & 1680 & 550 & 0.0080 & -0.38 & 0.17 \\
23\tablenotemark{a} & 590 & 460 & 0.0091 & -0.30 & 0.09 \\
25 & 720 & 170 & 0.0190 & 0.03 & 0.11 \\
26 & 400 & 1 & 0.0200 & 0.06 &  0.00 \\
27 & 530 & 63 & 0.0300 & 0.25 & 0.11 \\
29\tablenotemark{a} & 430 & 61 & 0.0107 & -0.23 & 0.09 \\
30\tablenotemark{a} & 360 & 42 & 0.0151 & -0.09 & 0.16 \\
31\tablenotemark{a} & 220 & 34 & 0.0111 & -0.23 & 0.13 \\
33 & 230 & 49 & 0.0083 & -0.34 & 0.11 \\
36\tablenotemark{a} & 120 & +120 & 0.0119 & -0.21 & 0.17 \\
37 & 5000 & 3990 & 0.0033 & -0.80 & 0.20 \\
38 & 160 & 16 & 0.0400 & 0.40 & 0.024 \\
39 & 220 & 160 & 0.0109 & -0.23 & 0.14 \\
40 & 480 & 130 & 0.0103 & -0.25 & 0.09 \\
42 & 770 & 420 & 0.0185 & -0.06 & 0.36 \\
43 & 350 & 180 & 0.0072 & -0.47 & 0.27 \\
44 & 200 & 53 & 0.0228 & 0.09 & 0.22 \\
46 & 460 & 130 & 0.0095 & -0.28 & 0.05 \\
47 &  100 & +100 & 0.0081 & -0.38 & 0.21 \\
\hline
Mask 2 & & & & & \\
\hline
5* & 330 & 25 & 0.0243 & 0.15 & 0.10 \\
10* & 300 & 64 & 0.0393 & 0.39 & 0.06 \\
14* & 9100 & 620 & 0.0008 & -1.37 & 0.13 \\
17* & 530 & 440 & 0.0168 & -0.28 & 0.60 \\
18* & 240 & 64 & 0.0190 & -0.06 & 0.30 \\
26* & 520 & 42 & 0.0364 & 0.34 & 0.11 \\
28* & 2000 & 1540 & 0.0047 & -0.62 & 0.18 \\
29* & 400 & 3 & 0.0400 & 0.40 & 0.00 \\
34* & 120 & 24 & 0.0351 & 0.32 & 0.12 \\
36* & 440 & 320 & 0.0272 & 0.20 & 0.09 \\
\enddata
\tablenotetext{a}{Also on mask 2.}
\label{agemetalTable}
\end{deluxetable}

%% file: table5.tex
\begin{deluxetable}{rcc}
\tablecolumns{3}
\tablewidth{0pc}
\tablecaption{Coefficients for $v_{\text{cluster}}$ vs. $R_{\text{semi-minor}}$ Fits}
\tablehead{\colhead{Data Set} & \colhead{Intercept} & \colhead{Slope}}
\startdata
GCs & 207 $\pm7$ & -5 $\pm5$ \\
YMCs & 249 $\pm8$ & 16 $\pm3$ \\
HI gas & 243 $\pm3$ & 18 $\pm1$ \\
\enddata
\label{velradSlopesTable}
\end{deluxetable}

%% file: table6.tex
\begin{deluxetable*}{lcccc}
\tablecolumns{5}
\tablewidth{0pc}
\tablecaption{Rotational Velocity and Velocity Dispersion Comparisons}
\tablehead{\colhead{Data Set} & \colhead{$v_{\text{rot}}$~(km/s)} & \colhead{$v_{\text{rot}}$ Error\tablenotemark{a}~(km/s)} & \colhead{$\sigma$~(km/s)} & \colhead{$v_{\text{rot}}/\sigma$}}
\startdata
{\bf MW\tablenotemark{b}} & & & & \\
Young Clusters & 215 &  & 10 & 22 \\
Old Open Clusters & 211 & 7 & 28 & 7.5 \\
Disk/bulge GCs & 156-193 &  & 67 & 2.3 \\
Halo GCs & 50 & 23 & 114 & $0.4\pm0.2$ \\
 & & & & \\
{\bf M33\tablenotemark{c}} & & & & \\
Young Clusters & 87 & 11 & $17$ & $5.1_{-1.1}^{+1.4}$\\
Disk/bulge GCs & -2 & 51 & $54$ & $0.04^{+1.02}$ \\
Halo GCs & 7 & 82 & $82$ & $0.09^{+1.2}$ \\
& & & & \\
{\bf All M101} & & & & \\
HI gas & 208 & 61 & 17 & $12_{-6}^{+5}$ \\
YMCs & 228 & 116 & 25 & $9_{-4}^{+5}$ \\
GCs & & & 66 & \\
 & & & & \\
{\bf M101, $r_{\text{gc}} > 5$~kpc} & & & & \\
HI gas & 204 & 17 & 12 & $18_{-1}^{+2}$ \\
YMCs & 229 & 99 & 25 & $9_{-4}^{+4}$ \\
\enddata
\tablenotetext{a}{For M101 data, the error is the standard deviation of the individual cluster or gas $v_{\text{rot}}$.}
\tablenotetext{b}{From \citet{lyn87}, \citet{sco95}, \citet{arm89}, \citet{cot99}, and \citet{zin85}.}
\tablenotetext{c}{From \citet{cha02}.  For M33, disk/bulge GCs are within 2.25~kpc while halo GCs are outside 2.25~kpc.}
\label{vrotTable}
\end{deluxetable*}